\documentclass[fleqn,usenatbib]{mnras}

\usepackage{newtxtext,newtxmath}
\usepackage[T1]{fontenc}

\DeclareRobustCommand{\VAN}[3]{#2}
\let\VANthebibliography\thebibliography
\def\thebibliography{\DeclareRobustCommand{\VAN}[3]{##3}\VANthebibliography}

\usepackage{graphicx}	
\usepackage{amsmath}	
\usepackage[english]{babel} 
\usepackage[usenames]{color} 
\usepackage[normalem]{ulem}

\newcommand{\kmps}{\rm km~s\ensuremath{^{-1} }\,}

\newcommand{\Msun}{\rm M\ensuremath{_\odot}}
\newcommand{\Msunyr}{\rm M\ensuremath{_\odot}~yr\ensuremath{^{-1}}\,}
\newcommand{\Msunpc}{\rm M\ensuremath{_\odot}~pc\ensuremath{^{-2} }\,}

\newcommand{\Oo}{\displaystyle}
\newcommand{\aH}{\ensuremath{\rm [\alpha/H]}\,}
\newcommand{\aFe}{\ensuremath{\rm [\alpha/Fe]}\,}

\newcommand{\FeH}{\ensuremath{\rm [Fe/H]}\,}

\newcommand{\Rini}{\ensuremath{\rm \langle R_{ini} \rangle}\,}
\newcommand{\Rfin}{\ensuremath{\rm \langle R_{fin} \rangle}\,}

\begin{document}

\title[Bimodality of \aFe-\FeH distributions in Milky Way-type galaxies]{Bimodality of \aFe-\FeH distributions is a natural outcome of dissipative collapse and disc growth in Milky Way-type galaxies}

\author[S. Khoperskov et al.]{Sergey Khoperskov$^{1,2,3}$\thanks{E-mail: sergey.khoperskov@gmail.com}, Misha Haywood$^{4,5}$, Owain Snaith$^{4}$, Paola Di Matteo$^{4,5}$, \newauthor Matthew Lehnert$^{5}$, Evgenii Vasiliev$^{3,6,7}$, Sergey Naroenkov$^{3}$, Peter Berczik$^{8,9,10}$ \\ $^{1}$Leibniz Institut f\"{u}r Astrophysik Potsdam (AIP), An der Sternwarte 16, D-14482, Potsdam, Germany\\ $^2$Max-Planck-Institut f\"{u}r extraterrestrische Physik, Gie{\ss}enbachstrasse 1, 85748 Garching, Germany \\ $^3$Institute of Astronomy, Russian Academy of Sciences, 48 Pyatnitskya St., Moscow, 119017, Russia \\ $^{4}$GEPI, Observatoire de Paris, PSL Universit{\'e}, CNRS,  5 Place Jules Janssen, 92190 Meudon, France \\ $^{5}$Sorbonne Universit\'{e}, CNRS UMR 7095, Institut d'Astrophysique de Paris, 98bis bd Arago, 75014 Paris, France,  \\ $^6$Southern Federal University, Rostov on Don 344090, Russia\\ $^7$Lebedev Physical Institute, Russian Academy of Sciences, 53 Leninsky Ave., 119991, Moscow, Russia \\ $^8$National Astronomical Observatories and Key Laboratory of Computational Astrophysics, CAS, 20A Datun Rd., Chaoyang District, Beijing 100101, China \\ $^9$Astronomisches Rechen-Institut am Zentrum fuer Astronomie der Universitaet Heidelberg, Moenchhofstrasse 12-14, 69120 Heidelberg, Germany \\ $^{10}$Main Astronomical Observatory, National Academy of Sciences of Ukraine, MAO/NASU, 27 Akad. Zabolotnoho St 03680 Kyiv, Ukraine}

\maketitle

\begin{abstract}
We present a set of self-consistent chemo-dynamical simulations of Milky Way-type galaxies formation and evolution to study the origin of the bimodality of $\alpha$-elements in stellar populations. We explore how the bimodality is related to the geometrically and kinematically defined stellar discs, gas accretion and stellar radial migration. We find that the two $\alpha$-sequences are formed in quite different physical environments. The high-$\alpha$ sequence is formed early from a burst of star formation in a turbulent, compact gaseous disc which forms a thick disc. The low-$\alpha$ stellar population is the result of quiescent star formation supported by the slow accretion of enriched gas onto a radially extended thin disc. Our simulations suggest that stellar feedback-driven outflows during the formation of the thick disc are responsible for the enrichment of the surrounding gaseous halo, which subsequently feeds the disc on a longer time-scale. During the thin disc phase, chemical evolution reaches an equilibrium metallicity and abundance, where the stars pile-up. This equilibrium metallicity decreases towards the outer disc, generating the ridge line that forms the low-$\alpha$ sequence. We identify a second mechanism capable of creating a low-$\alpha$ sequence in one of our simulations. A rapid shutdown of the star formation due to feedback at the end of the thick disc phase, suppresses the chemical enrichment of the halo gas, which, once accreted onto the star-forming disc, dilutes the ISM at the beginning of the thin disc formation. Both mechanisms can operate in a galaxy, but the former is expected to occur when star formation efficiency ceases to be dominated by the formation of the thick disc, while the latter can occur in the inner regions. Being the result of the presence of low and high gas density environments in a galaxy, the bimodality is independent of any particular merger history, suggesting that it could be much more widespread than has been claimed. We also find that radial migration has a negligible effect on the \aFe-\FeH distribution over time, suggesting that $\alpha$-bimodality results purely from the presence of different star formation regimes over the galaxy's formation. 
\end{abstract}

\begin{keywords}
galaxies: formation -- galaxies: evolution  -- Galaxy: formation -- Galaxy: evolution -- Galaxy: abundances -- Galaxy: disc
\end{keywords}

\section{Introduction}

The Milky Way is a typical disc galaxy composed of thin and thick stellar discs, a central pseudo-bulge and an old stellar halo. The latter seems to be dominated by a single major merger at early epoch~\citep{2018MNRAS.478..611B,2018ApJ...863..113H,2018Natur.563...85H}. For a long time the bulge has been seen as an old classical spheroid, but models and observations now show that most of the mass is concentrated in a boxy-peanut pseudo bulge made of disc(s) stars which passed through the phase of secular evolution driven by the buckling instability of the bar~\citep{2010ApJ...720L..72S, 2011MNRAS.416L..60B, 2015A&A...577A...1D,2017MNRAS.467L..46A,2017MNRAS.469.1587D,2018A&A...616A.180F,2019A&A...628A..11D}.

The Milky Way stellar disc is a complex structure containing thick and thin components, a bar and spiral arms. The thick disc  was recognized as a component of the Milky Way distinct from the thin disc by~\cite{1983MNRAS.202.1025G}. Thick disc stars are significantly older than those of the thin disc, which means that they contain a fossil record of the formation processes of the early phases of the Milky Way evolution. However, parsing of thick/thin disc stars is quite problematic because they intersect much in phase space due to several dynamical effects, such as radial migration~\citep{2002MNRAS.336..785S}, the presence of spiral arms and the bar evolution~\citep{2000AJ....119..800D, 2005AJ....130..576Q, 2009ApJ...700L..78A, 2011A&A...527A.147M} and gravitational interactions with the environment~\citep{2006ApJ...641L..33W,2010ApJ...709.1138D,2011Natur.477..301P,2013MNRAS.429..159G, 2017MNRAS.465.3446G}. In such a context, the dissection of thin/thick discs by chemical composition is the preferred tool. Moreover, the chemical composition of its stars is related to the star formation history of a galaxy \citep{2015A&A...578A..87S}, and therefore to the evolution of the gas content during galactic evolution. 

One of the most striking chemical patterns discovered in the Milky Way disc is the two sequences visible in the abundance ratios of alpha elements as a function of metallicity for stars in the solar vicinity~\citep{1996ASPC...92..307G, 1998A&A...338..161F, 2004AN....325....3F,2006MNRAS.367.1329R,2011A&A...535L..11A,2014A&A...562A..71B,2012A&A...545A..32A,2020MNRAS.tmp..333H}. Over the last few years a number of studies exposed the clear separation of $\alpha$-abundance for a given metallicity on larger spatial scales~\citep{2014A&A...564A.115A, 2015ApJ...808..132H,2019arXiv191209778Q}, together with the prevalence of the high-$\alpha$ sequence in the inner Galaxy (R$<$6~kpc) and of the low-$\alpha$ sequence in the outer disc (R$>$10~kpc). The existence of a distinct $\aFe$-bimodality across the Milky Way disc suggests two different formation paths for high- and low-$\alpha$ sequences but the mechanism by which this bimodal distribution appeared remains debated. 
Since it is traditionally accepted that high- and low-$\alpha$ sequences correspond to the thick and thin Galactic disc~\citep[see., e.g.,][]{1998A&A...338..161F,2013A&A...560A.109H}, respectively, the problem of understanding the global chemical abundance patterns is ultimately linked to the understanding of the formation scenarios of the Milky Way disc components.

Before going further, the ambiguity between two types of bimodality in the \FeH-\aFe distribution must be clarified. Bimodality (1), whose origin is addressed here, is the presence of two clearly defined  \aFe sequences in this plane, and it is visible at all \FeH of disc stars. This bimodality is evident even when not taking into account the density of stars along the two sequences. Bimodality (2) reveals itself only in the inner disc, with a bimodal distribution of \aFe abundances. This second bimodality (2) comes from the quenching phase that occurred between the formation of the thick disc and the inner thin disc~\citep[see][]{2016A&A...589A..66H}. The two bimodalities are generated by different processes: the first bimodality (the existence of two \aFe sequences), cannot be generated by a simple gap in the star formation history.

\begin{figure}
\includegraphics[width=0.9\hsize]{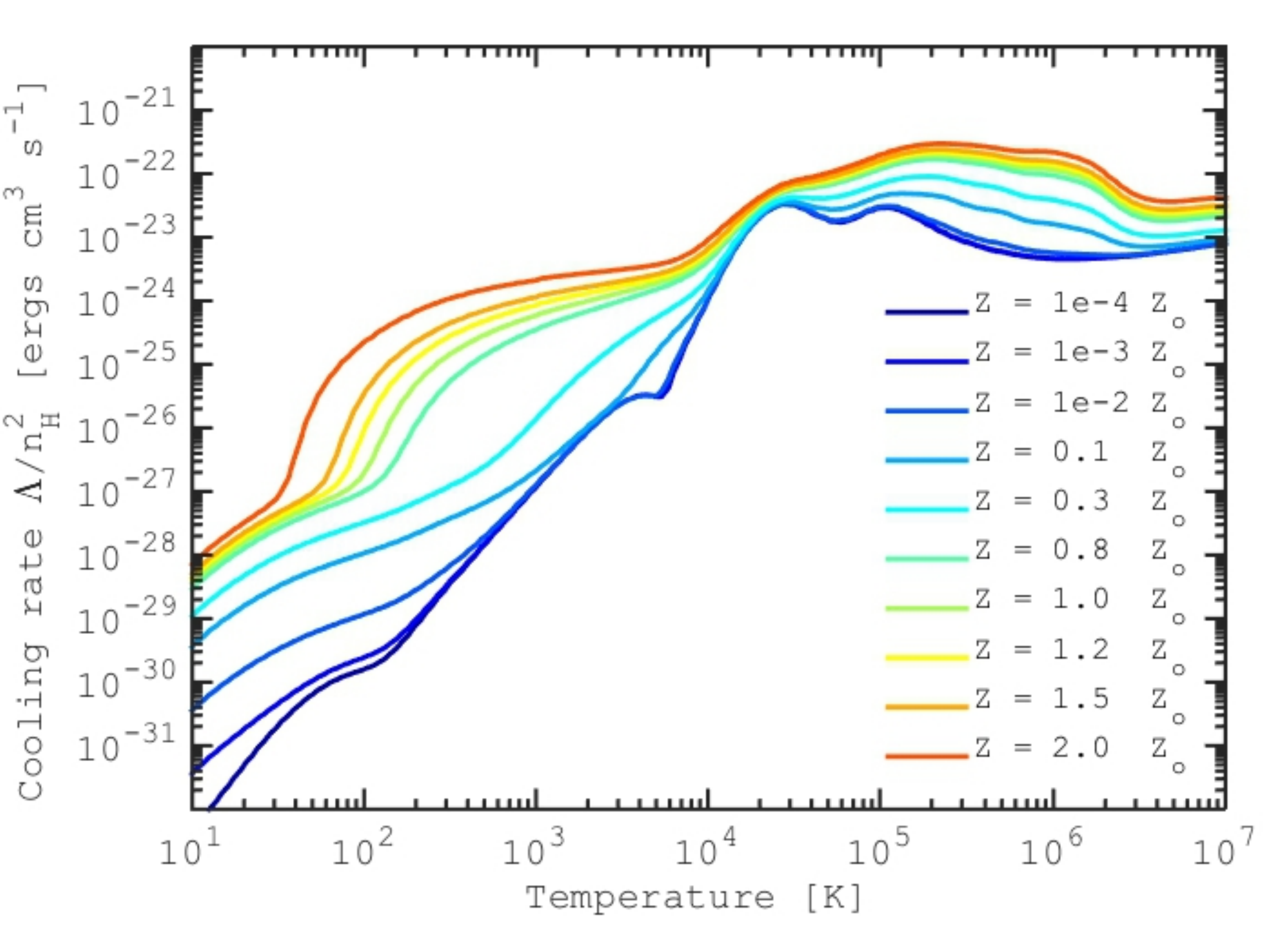}
\caption{Cooling functions adopted for the calculation of $\Lambda/n^2$~(ergs~cm$^3$~s$^{-1}$) for different values of gas total metallicity~($\rm Z$).}\label{fig::cooling}
\end{figure}

Several studies have tried to reproduce the observed \aFe-bimodality~(1), proposing radically different solutions. For instance, \cite{2017MNRAS.472.3637G} tried to reproduce the bimodality as it is observed in the solar vicinity with a hiatus in the chemical track in the $\alpha$-metallicity plane generated by a gap between two sequential gas infalls~\citep[see also][]{1997ApJ...477..765C, 2009IAUS..254..191C,2019A&A...623A..60S}. \cite{2013A&A...560A.109H} advocated that the inner and outer discs, dominant in the high and low-$\alpha$ sequences respectively, both present at the solar vicinity, cannot be represented by the same evolution, and for this reason they were described by two different models in \cite{2015A&A...578A..87S}. It was also claimed by these authors that sustained star formation activity must have formed the thick disc, as was confirmed by the measurement of the star formation history in \cite{2014ApJ...781L..31S}, the chemical homogeneity of this population at all ages implying a high level of turbulence in the gas \citep[see also][]{2015A&A...579A...5H}. This picture was completed more recently in \cite{2019A&A...625A.105H}, where they suggested that while the high-$\alpha$ sequence is a temporal sequence, the low-$\alpha$ sequence is made of parallel evolution occurring at different radii, starting at $R\gtrsim6$~kpc. 

\cite{2004ApJ...612..894B,2005ApJ...630..298B} showed that the thin and thick disc stars can have parallel tracks in $\aFe-\FeH$ plane in simulations with a very high star formation rates triggered by the early gas rich merging epoch. More recent simulations emphasized the possible importance of external processes in the formation of the dichotomy in the \aFe distribution. By using Auriga cosmological simulations \cite{2018MNRAS.474.3629G} found only one galaxy with a clear \aFe-bimodality in the outer disc, produced when the gaseous disc shrinks after the high-\aFe sequence forms, causing a decrease in SFR while the gas transitions to a low-\aFe state. Subsequent low-metallicity gas accretion grows the low-\aFe sequence in an inside-out fashion at lower metallicities compare to the end of the high-\aFe sequence. By analysing 133 simulated EAGLE Milky Way-mass galaxies, \cite{2018MNRAS.477.5072M} found that  a prominent disc bimodality is rare. They deduce from their analysis that a bimodality is established only when  an early, first, rapid phase of gas accretion is followed by a slower, usually later, second phase, where the gas is mainly acquired from a galaxy merger. In an alternative approach both high- and low-$\alpha$ sequences can be formed in parallel~(or simultaneously). \cite{2019MNRAS.484.3476C} presented a galaxy formation simulation where gas clumps self-enrich in $\alpha$-elements due to their high star formation rate density and produce the high-$\alpha$ sequence while the low-$\alpha$ sequence is produced contemporaneously by the star formation in a smooth surrounding disc, implying a temporal overlap in the formation of the two $\alpha$-sequences. One of the most recent chemodynamical study of the \aFe bimodality in simulated galaxies has been made by~\cite{2020MNRAS.491.5435B} who conducted an in-depth analysis of chemical patterns in Milky Way-type galaxies from the NIHAO-UHD project. In the four simulation analyzed in their paper, \cite{2020MNRAS.491.5435B} finds that the high-$\alpha$ sequence is formed first, and the low-$\alpha$ sequence is a generic consequence of a gas-rich merger at some point in the galaxy's evolution. More recently  \cite{2020arXiv200606012R,2020arXiv200606011R} and \cite{2020arXiv200606008A} presented a zoom-in simulation of the Milky Way analogue where cosmological accretion leads to the rapid formation of an outer, metal-poor, low-\aFe disc around the inner, metal-rich galaxy containing the old high-\aFe stellar populations. In this scenario, the galaxy is fueled by two distinct gas flows which leads to a bimodality in \aFe over a range of \FeH. Importance of the outflow-to-infall ratio of metals was also noticed by ~\cite{2020MNRAS.496...80V} who found an \aFe-\FeH dichotomy in a Milky Way-mass galaxy with quiet formation history.

\begin{figure*}
\begin{center}
\includegraphics[width=0.8\hsize]{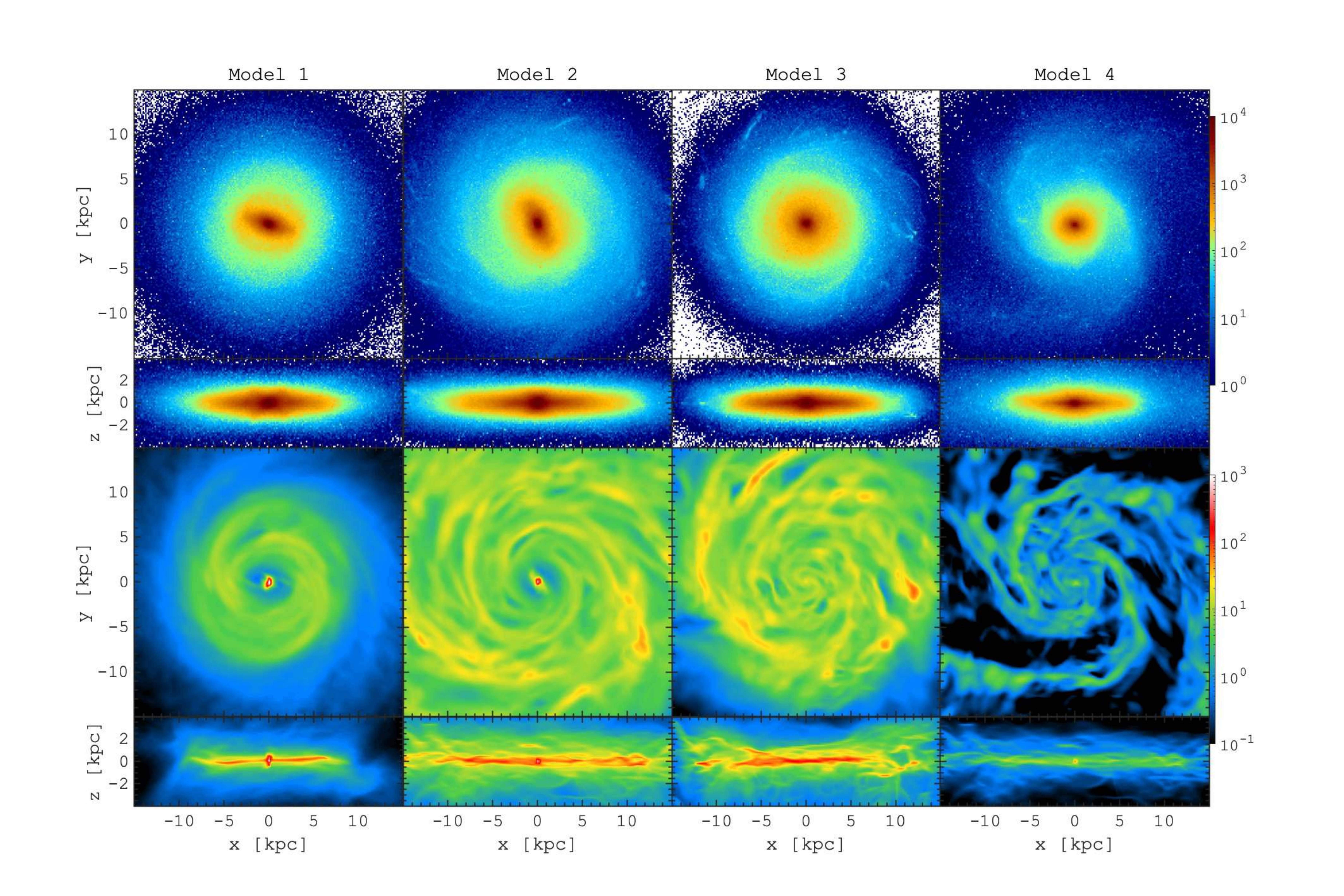}
\caption{Face-on and edge-on density maps~(\Msunpc in log-scale) of the stellar~(top frames) and gaseous~(bottom frames) components for the four simulated galaxies after $10$~Gyr of evolution.}\label{fig::maps_at_end}
\end{center}
\end{figure*}

In this paper, we present a set of idealized galaxy formation models via cooling of gas within an evolving galactic dark matter halo. Our work aims to explore some basic chemodynamical processes that can lead to the formation of the \aFe-bimodality in simulated Milky Way-type disc galaxies. We analyze isolated galaxy formation simulations where the gas is able to feed the disc, driving the evolution of chemical abundances of the different galactic components.   The choice to study, in particular, galaxies that evolve in isolation is dictated by the desire to understand if and under what conditions a \aFe bimodality can be generated in a disc galaxy, even in the absence of mergers.
The paper is structured as follows. In Section~\ref{sec::models}, we present our numerical methods, models setup and briefly summarise the global evolution of simulated galaxies.  In Section~\ref{sec::gas} we explore the enrichment of the multi-phase gaseous component feeding the galaxies over time. In Section \ref{sec::results} we explore the pathways to the \aFe bimodality as well as the spatial, kinematical and age structure of the stellar populations. In Section~\ref{sec::migration} we describe the impact of stellar radial migration on the observed chemical abundance patterns across both thin and thick discs. Finally, the results are discussed and summarized in Sections~\ref{sec::discuss} and \ref{sec::conclusions}, respectively.

\begin{figure*}
\begin{center}
\includegraphics[width=0.8\hsize]{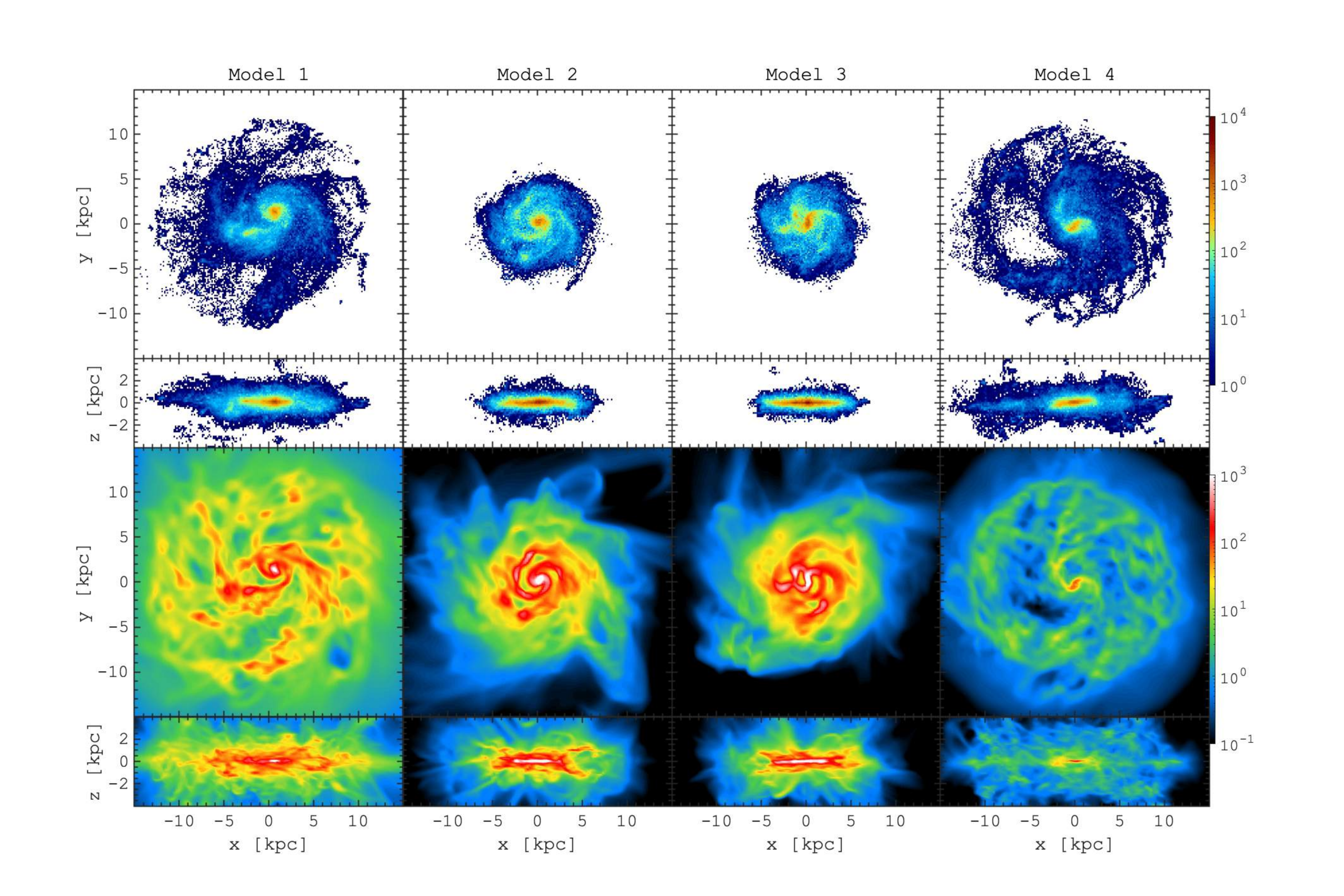}
\caption{Same as in Fig.~\ref{fig::maps_at_end} but at early times of evolution, $t=500$~Myr. Our simulated galaxies went through a clumpy phase with rapid internal evolution, similar to that of progenitors of Milky Way-type galaxies at high redshifts. Another important characteristic of the early evolution is the gaseous outflows caused by the stellar feedback~(best visible in models 2 and 3 with a compact disc).}\label{fig::chains}
\end{center}
\end{figure*}

\section{Simulation set up}\label{sec::models}
This section provides an overview of the simulations and their subgrid physics routines. 

\subsection{Models setup}

Our models consist of a disc galaxy forming inside a hot corona of pressure-supported gas embedded in a dark matter halo, an approach that has been used extensively for studying isolated disc galaxy formation and evolution~\citep[see, e.g.,][]{2014MNRAS.445.3352C, 2016MNRAS.459.3326A,2019MNRAS.484.3476C} and simulations of mergers~\citep{2016ApJ...821...90A}. We believe that this approach may also be relevant specifically to the Milky Way formation/evolution which did not experience any significant mergers over the last $\approx 9$~Gyr~\citep{2018ApJ...863..113H,2018Natur.563...85H,2019A&A...632A...4D,2020MNRAS.498.2472K} implying that the Milky Way stellar mass has grown mostly by gas accretion and in-situ star formation~\citep{2015A&A...578A..87S,2019A&A...625A.105H}. 

In this paper we consider four models where the initial gaseous halo has different scale length, different initial mass and different $\lambda$ spin parameters. These parameters are listed in Table~\ref{tab::tab}. The gas is initially a Plummer sphere density distribution with a temperature profile to yield an approximate hydrostatic equilibrium~\citep[see, e.g.,][]{2013MNRAS.428.1055A}. The angular momentum follows a radial profile of $j \propto r$, where $j$ is the specific angular momentum~\citep[see, also,][]{2008ApJ...675L..65R,2014MNRAS.445.3352C,2019MNRAS.484.3476C} which gives us a spin values consistent with those obtained from collisionless cosmological simulations~\citep{2001ApJ...555..240B}. Due to the large uncertainties in the actual gas metallicity at high redshift, we rely on the assumption of a primordial chemical composition of the gas, with a 24\% mass fraction of helium, and the rest of the mass in hydrogen. No star particles are present at the beginning of the simulations since all stars form out of gas that cools and reaches conditions to trigger star formation~(see below). The initial state for the rotating gaseous halo embedded into the live DM halo has been generated using the iterative method described in~\cite{2009MNRAS.392..904R}. The dark matter halo is set up using a population of $10^6$ collisionless particles following a Plummer distribution with a scale of $10$~kpc and the spin values adopted as for the gas. We truncate the dark matter halo at $160$~kpc. The particle mass is $2.57\times10^5$~\Msun, and the gravitational softening length is set for DM to $250$~pc. 
 
The simulations were evolved with the $N$-body+Total Variation Diminishing hydrodynamical code~\citep{2014JPhCS.510a2011K}. For the $N$-body system integration and gas self-gravity, we used our parallel version of the TREE-GRAPE code~\citep[][]{2005PASJ...57.1009F} with multithread usage under the SSE and AVX instructions. In recent years we already used and extensively tested our hardware-accelerator-based gravity calculation routine in several galaxy dynamics studies where we obtained accurate results with a good performance~\citep{2016MNRAS.462.3727P,2017MNRAS.470...20S,2017A&A...604A..75M,2018A&A...611L...2K,2018A&A...620A.154K,2018MNRAS.481.3534S,2019A&A...622L...6K}. For the time integration, we used a leapfrog integrator with a fixed step size of $0.1$~Myr. In the simulation we adopted the standard opening angle $\theta = 0.7$. Although we do not force a constant value of the initial mass of newborn star particles, the vast majority~($>90\%$) of them have a mass of $\approx 3.8\times10^3$~\Msun. This allows us to use a constant value of gravitational softening parameter of $50$~pc for star particles. The number of stellar particles at the end of simulations for different models vary in the range of $\approx (3-5)\times10^6$ which makes our galaxies more detailed compared to similar studies based on large-scale cosmological simulations~\citep[e.g.,][]{2018MNRAS.477.5072M}. Dynamics of the ISM is simulated on a Cartesian grid with static mesh refinement and a minimum cell size of $50$~pc close to the galactic disc plane. The hydrodynamical part also includes radiative cooling~(see next Section) and heating terms and idealized magnetic field treatment with constrained transport approach~\citep{2014JPhCS.510a2011K}. 

\subsection{Subgrid physics}
In our simulations a gaseous cell undergoes star formation if : i) the gas mass is $> 2\times10^5$~\Msun, (ii) the temperature $T$ is lower than $100$~K and (iii) is part of a converging flow. The efficiency of star formation is set to $0.05$, i.e. 5\% of the gas eligible to form a new star particle per dynamical time. There is no direct dependency of the star formation efficiency on the gas metallicity. Newborn stellar particles inherit the velocity and elemental abundances of their parent gas cells. Although in our models we allow the formation of new particles in a wide range of masses, we found that ~95-98\% of particles have the mass of about $4 \times 10^4$ which does not contribute significantly to artificial heating due to constant softening. The mass of newborn stellar particles is not fixed in our simulations, however about $95-98\%$ of particles have the mass of about $4 \times 10^4$, so each can be considered as a single stellar population. We assume the initial distribution of stellar masses~(IMF) to be described by the parameters from~\cite{2001MNRAS.322..231K}, normalized between $0.1$~and $100$~\Msun. At each time step, we calculate the amount of gas returned, the mass of the various species of metals, the number of SNII or SNIa for a given initial mass and metallicity, the cumulative yield of various chemical elements~(O, Mg, Si, Fe, H), the total metallicity, and the total gas released. Released mass of different elements are treated separately as passive scalars which obey non-homogeneous transport equations and thus are advected with the gas~\citep[see, e.g.,][]{2012ApJS..198....7M}. Therefore we consider the ISM as a mixture of several species (H, He, Si, Mg, O, Fe, other metals) which is sufficient for modelling the galactic chemical evolution. Following standard approach for the passive scalars, these species are mixed within a given cell but abundances are allowed to be different in different cells. Following the chemical evolution models by~\cite{2015A&A...578A..87S}, at each time step, the mass and metals released from evolving stellar populations are transferred from stellar particles to their neighbours, with weights calculated using the initial, rather than current, mass of the particle. Feedback associated with the evolution of massive stars is implemented as an injection of thermal energy in a nearby gas cell proportional to the number of SNII, SNI and AGB stars. Outflows develop naturally without the need to specify an initial gas velocity and do not require that radiative cooling or hydrodynamic forces to be temporarily disabled~\citep[see, also,][]{2016ApJ...833..202K}. Such an approach was successfully used in a number of our previous studies~\citep{2016MNRAS.455.1782K,2017MNRAS.468..920K,2018A&A...609A..60K} and provides a realistic ISM treatment and Kennicutt-Schmidt-type relations for both low- and high- star formation regimes of galaxy evolution.

In the energy equation for hydrodynamics, we take into account cooling processes using the tabulated non-equilibrium cooling curves which depend on the metallicity of the gas~(see Fig.~\ref{fig::cooling}). The full description of our method of cooling rate calculations and the references to the atomic data can be found in \citet{2011MNRAS.414.3145V,2013MNRAS.431..638V}.  Although we do not consider the detailed chemical reactions network in the ISM, a tabulated metallicity dependent cooling rates provide us a reasonable agreement of the ISM structure~(pressure-temperature relation, see Fig.~\ref{fig::ISM}) with widely-used state-of-the-art gravito-hydrodynamics codes~\citep{2016ApJ...833..202K}.

In brief, the non-equilibrium calculation includes the ionization kinetics of all ionization states for the following chemical elements H, He, C, N, O, Ne, Mg, Si, Fe as well as molecular hydrogen kinetics at $T<10^4$~K. We take into account the following major processes in a collisional gas: collisional ionization, radiative and dielectronic recombination as well as charge transfer in collisions with hydrogen and helium atoms and ions. In a low-temperature ($T\le 10^4$~K) gas the above-listed ionization states of the elements are supplemented by a standard set of species: H$^-$, H$_2$, H$_2^+$, D, D$^+$, D$^-$, HD, needed to model the H$_2$/HD gas-phase kinetics \citep{1997NewA....2..181A,1998A&A...335..403G}. Note that the H$_2$/HD cooling is efficient in low metallicity gas, $[Z/H]< -3$. 

The cooling rates are obtained for a gas cooled down isochorically from $10^8$ down to $10$~K. We calculate a grid of cooling rates covering the metallicity range from $[Z/H]= -4$ to $+0.3$. The grid is constructed so that the difference of cooling rate between neighbouring values of metallicity is smaller than a factor of two for any temperature. 

\subsection{Overview of the star formation histories of simulated galaxies}

As we described before, in this work we present the analysis of four Milky Way-type disc galaxies with stellar masses in the range of $(3.4-8.2)\times 10^{10}$~\Msun. Two galaxies have bars with boxy/peanut pseudo-bulges~(models 1 and 2) and the two other models present a spiral structure~(model 3 with a weak pattern and model 4 with a more prominent one) at the end of the simulation~(see Fig.~\ref{fig::maps_at_end}). This set of simulations, although being limited, allows us to follow the main paths of chemodynamical evolution of Milky Way-type disc galaxies. 

At the beginning of the simulations, the primordial gaseous haloes rapidly settle into thick disc-like structures. The turbulent gaseous phase at the beginning of our discs formation somehow resembles the gas-rich merger phase observed in  simulations with cosmological initial conditions~\citep[see, e.g.,][]{2004ApJ...612..894B}. In our models the high mass of the gaseous disc leads to its rapid fragmentation and to the formation of clumpy discs with an enhanced SFR. The chains of massive clumps~($10^8-10^9$~\Msun) dominate the structure of the galaxies, similarly to some previous simulations~\citep[][see Fig.~\ref{fig::chains}]{1999ApJ...514...77N, 2002MNRAS.330..707K, 2014ApJ...780...57B,2016MNRAS.456.2052I,2007ApJ...670..237B, 2019MNRAS.484.3476C,2020MNRAS.492.4716B} and the analytic calculations of the MW stellar structure at high-redshift~\citep{2011MNRAS.415.1280A,2019MNRAS.483...46Z}. We observe that after the fragmentation phase, some clumps fall to the centre, but some of them are dissolved, preventing the formation of massive bulges. Our galaxies pass through a stage of a rapid collapse of gas and intense star formation activity in a thick turbulent disc, where a high level of turbulence is driven by the stellar feedback, leading to an efficient mixing of metals in the disc. Despite a short time scale, this phase contributes significantly to the buildup of the stellar component. As can be seen in Fig.~\ref{fig::chains} the star formation rate density is sufficiently high to generate possible outflows~\citep{2014ApJ...789L..30L} that may have contributed to polluting the surrounding extended gaseous halo~\citep[see, e.g.,][]{2011MNRAS.416.1354D, 2016ApJ...824...57C,2017MNRAS.464.2796G}. At later epochs, the evolution is more quiescent, and settles to structures similar to those observed in nearby disc galaxies and in the Milky Way~(see Fig.~\ref{fig::maps_at_end}).

\begin{figure*}
\begin{center}
\includegraphics[width=0.3\hsize]{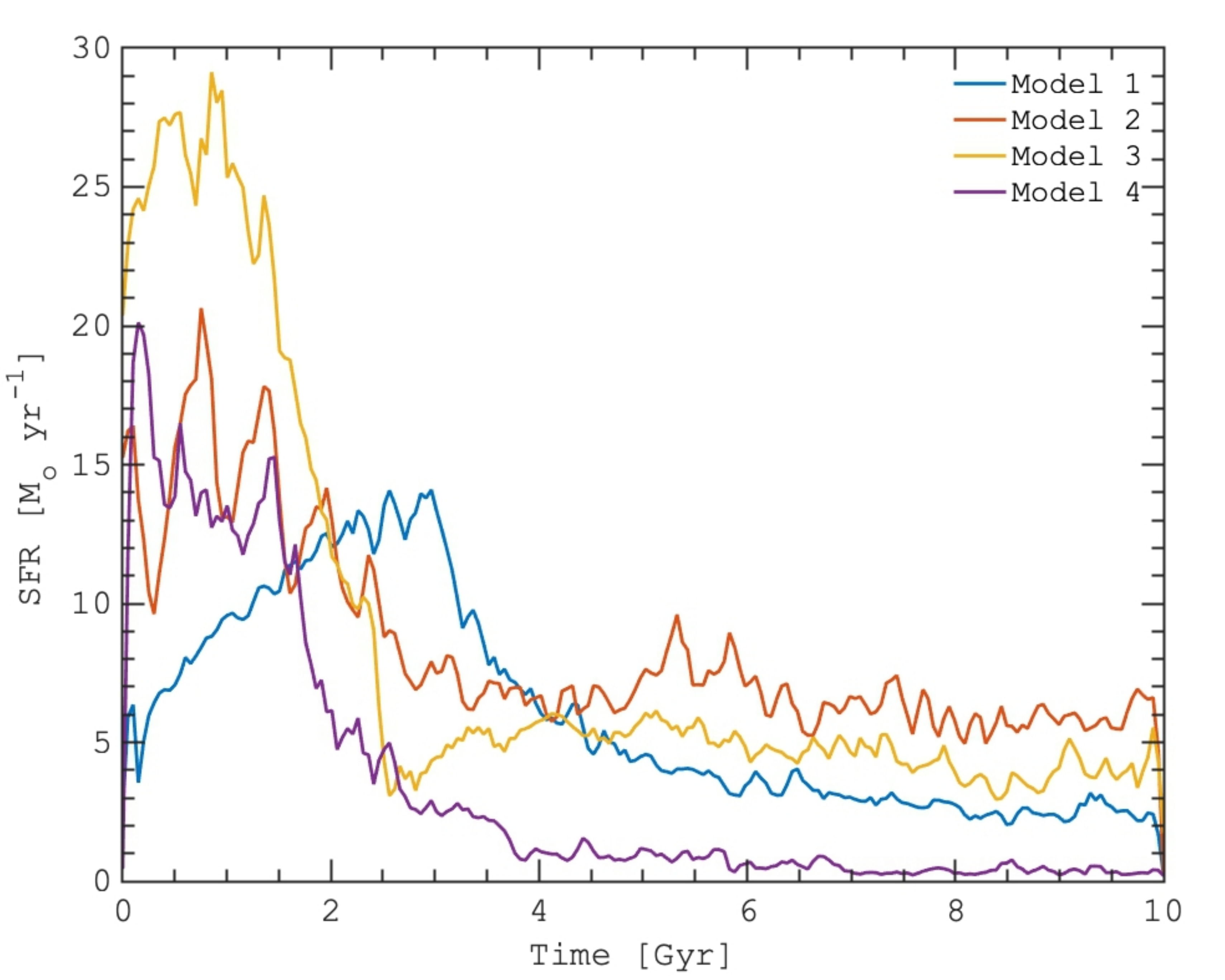}\includegraphics[width=0.3\hsize]{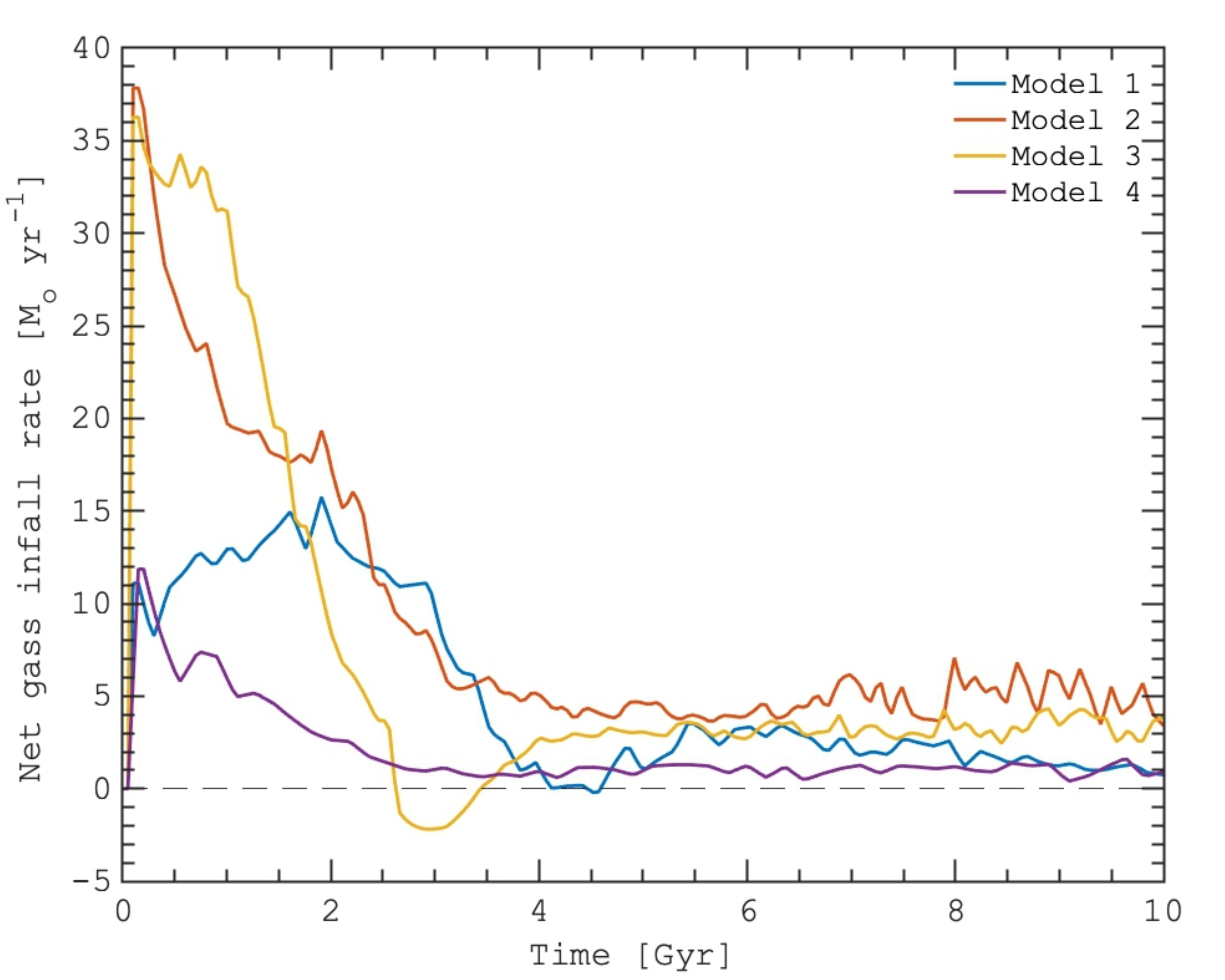}
\caption{{\it Left:} star formation history of simulated galaxies. {\it Right:} net gas infall rate as a function of time. At early times, accretion rate is high, reaching values up to $10-35$~\Msunyr and then decline towards net values within the range $0.5-5$~\Msunyr at the end of evolution. The sharp decrease of the gas infall rate at the end of the initial intense phase of star formation is caused by the intensive stellar feedback, which, e.g. in model 3 leads to the short period of net outflow of the gas.}\label{fig::sfr_infall}
\end{center}
\end{figure*}

To understand {\it the origin} of the diversity of chemical patterns, and of the $\alpha$-bimodality in particular, we need to explore the star formation, stellar feedback and the gas infall rates as a function of time. These are shown for the four simulations in Fig.~\ref{fig::sfr_infall}. Similar to ~\cite{2019MNRAS.482.3089N} we compute the gas accretion rates in units of solar masses per year by multiplying gas fluxes by their corresponding surface areas where the surface is defined by two parallel planes $\pm 1$~kpc from the disc plane. Of course the choice of the surface is arbitrary, but our selection provides a global view of the gas infall rate evolution. 
According to Fig.~\ref{fig::sfr_infall}, during a short initial phase~($1.5-3$~Gyr, depending on the model) primordial gas is rapidly accreted, giving rise to the formation of a significant fraction of the disc component. This episode is followed by a less intense~($0.5-5$~\Msunyr) new phase of gas accretion on much longer time-scales, resulting in the formation of a younger disc component. The star formation history~(see Fig.~\ref{fig::sfr_infall}) shows a similar temporal evolution to the gas accretion rate, which suggests that in our simulations the star formation rate is supported by the gas accretion rate~\citep[or re-accretion, see, e.g.,][]{2008MNRAS.386..935F, 2010MNRAS.404.1464M}. 
It seems to happen in two more or less distinct phases, namely, an early, initial intense period of star formation followed by a subsequent, lower star formation intensity phase. The SFH is also very much spatially segregated: the initial intense phase of star formation is limited to a few central kpcs, the early SFR intensity at 10 kpc being often less than 10\% its value in the inner regions of the discs. As we shall see in section~\ref{sec::results}, these spatial and temporal dichotomies, which are defined by the accretion history and distribution of the gas, are at the origin of the large-scale bimodality of the  \aFe-\FeH distribution.

\begin{figure*}
\begin{center}
\includegraphics[width=0.7\hsize]{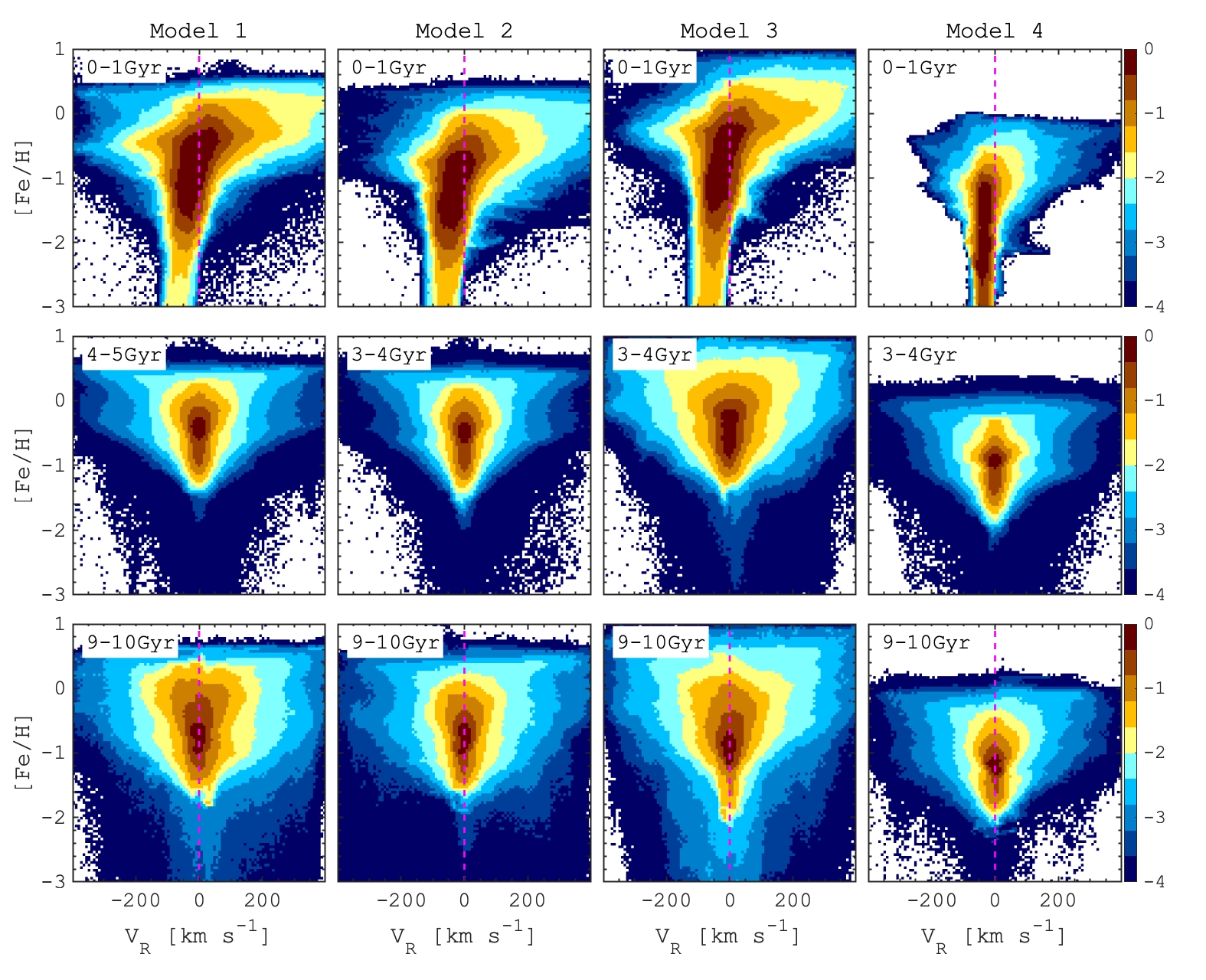}
\caption{Gas metallicity as a function of spherical radial velocity, with the vertical line demarcating inflow ($<0$) versus outflow ($>0$). For top frames we stack together 10 snaphots for the first $1$~Gyr of evolution. Second row represents the relation based on 10 snapshots after the end of the high-$\alpha$ sequence formation.  The bottom frames correspond to 10 snapshots at the end~($9-10$~Gyr) of the simulations. At early stages, during the burst of the SF, models 1-3 show that outflows have preferentially higher metallicity than inflows~\citep[see, also,][]{2020arXiv200907809P}, while in model 4 the outflows are less effective. At later stages of evolution we observe a balance between inflow/outflow of metals because of the low star formation activity in our galaxies.}\label{fig::enrich}
\end{center}
\end{figure*}

\begin{figure*}
\begin{center}
\includegraphics[width=0.8\hsize]{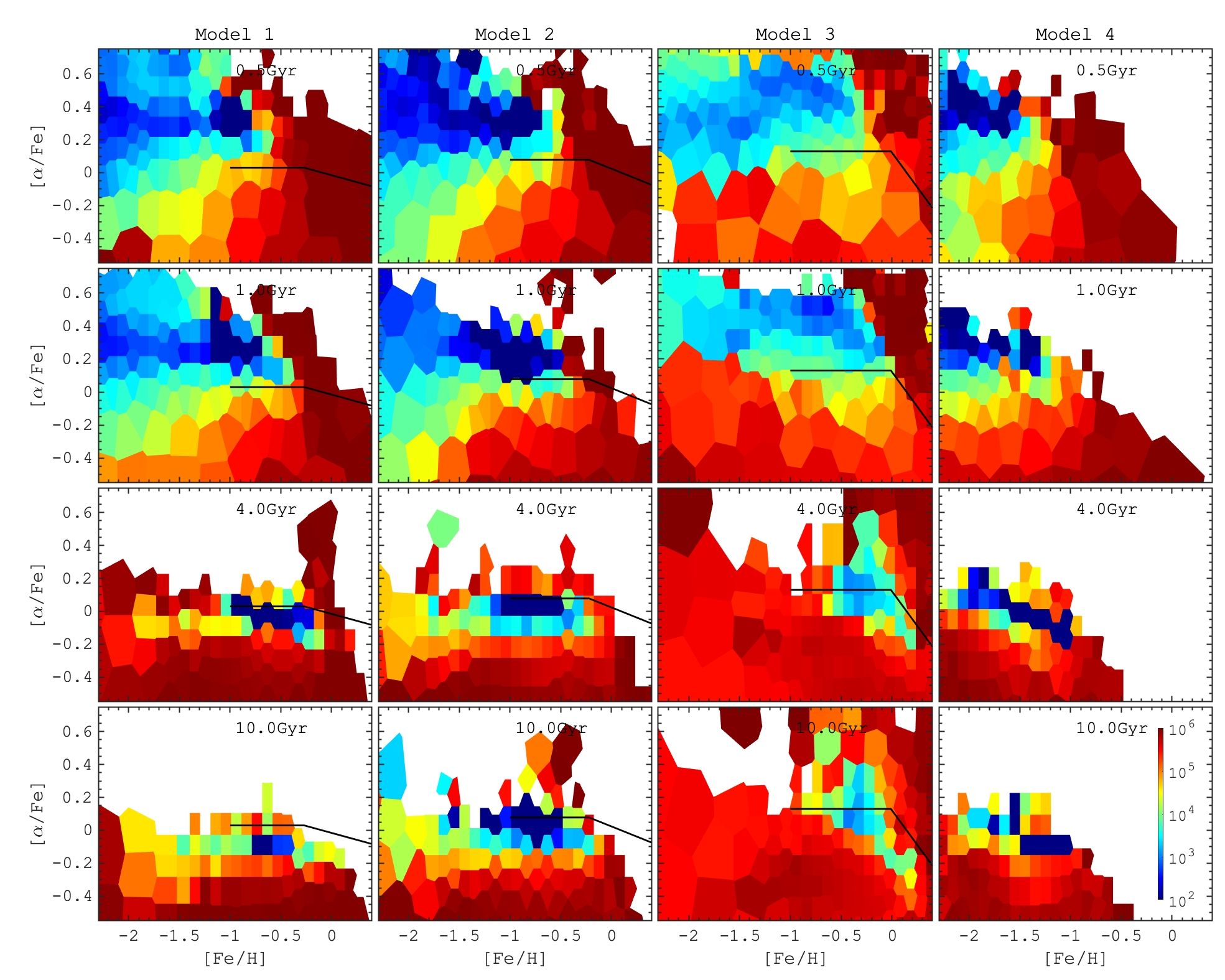}
\caption{Evolution of the median gas temperature distribution in $\aFe-\FeH$ plane. The temperature value is measured at snapshots marked in each frame ($0.5, 1, 4$ and 10~Gyr from top to bottom). The temperature scale is the same across the frames and it is shown in the bottom right frame. In our simulations star formation is allowed only in the cold ISM~($T<100$~K) which limits the inheritance of the chemical abundances of newly formed stars. Multi-phase structure of the ISM demonstrates that the amount of cold gas decreases with time, suggesting that thick discs form in the epoch of molecular gas domination. The distributions also illustrate the presence of, even at early times, metal-enriched hot gas which is not involved in the star formation activity due to long cooling time-scale. A large fraction of metals in the hot phase decreases the speed of the chemical evolution, e.g. compared to standard chemical evolution models. Black lines separate low- and high-$\alpha$ sequences in three models except model 4~(right column) where we do not find a prominent bimodality~(see Section~\ref{sec::results} for details).}\label{fig::gas_temp}
\end{center}
\end{figure*}

\begin{figure*}
\begin{center}
\includegraphics[width=0.8\hsize]{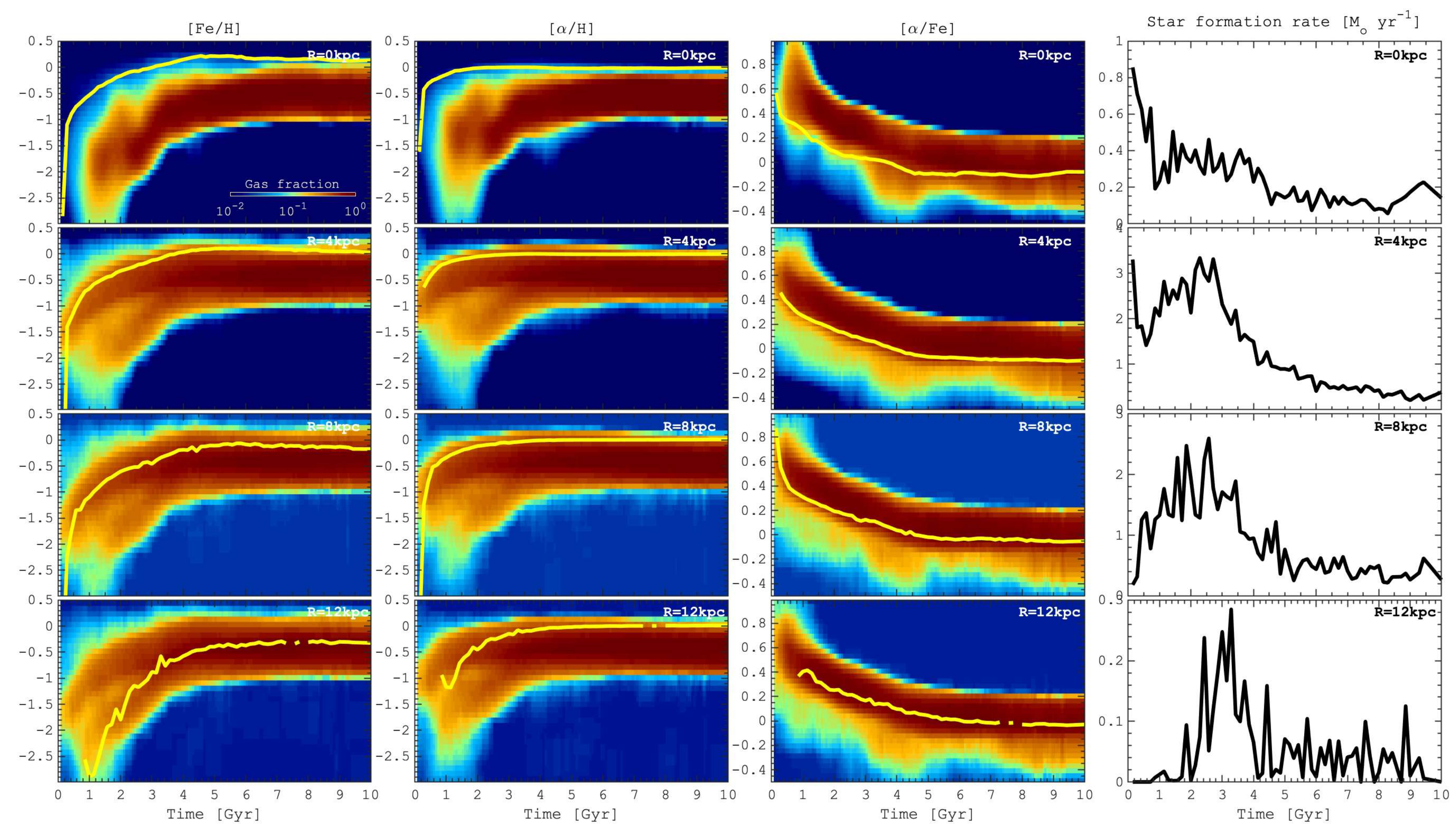}
\caption{Chemical evolution of stellar populations and the ISM enrichment history at different radii in model 1. From left to right: the average evolution of \FeH, \aH, \aFe and total star formation rate in $1$~kpc-width circular annuli: $0, 4, 8, 12$~kpc~(different rows). Chemical abundance evolution of new born stars~(age<10~Myr) is shown by yellow lines. Background colormaps represent a mass fraction~(normalized at each snapshot) of the gas according to the colour-bar plotted in the top left frame.}\label{fig::cem}
\end{center}
\end{figure*}

\section{Chemical enrichment of the multi-phase ISM}\label{sec::gas}

In chemical evolution models, the chemical evolution history is determined by both the star formation history, which determines the amount of metals that are rejected in the ISM as a function of time, and the ISM itself, with fixes the dilution of these elements. Both are controlled by the gas infall rate. As we have seen, the star formation history of our galaxies is critically determined by their infall history. However, in our simulations, the physics of the gas, described by a multi-phase ISM, introduces a more complex chemical cycle which is studied here.

In our simulations the source of the accreted gas is a pre-existing extended, initially hot ($T\sim10^6$ K), gaseous halo. The most intense phase of mass and metals exchange between the disc ISM and this surrounding circumgalactic medium~(CGM) is the thick disc formation at early epoch ($\approx 8-10$~Gyr ago), see Fig. \ref{fig::chains}.  During this phase and afterwards, the gas is heated through supernovae explosions, ejected in the halo through outflows, mixed by the induced turbulence in the disc. Figure~\ref{fig::gas_temp} illustrates the resulting complexity of the multi-phase structure of the ISM in $\aFe-\FeH$ plane at different times. To avoid some peculiarities in the chemical abundance evolution of individual elements we define $\rm [\alpha/H] = ([Si/H] + [Mg/H] + [O/H])/3$. Already at very early stages~($\approx0.5-1$~Gyr) we find the presence of gas with solar abundances, however, it is characterized by a very high temperature~($>10^5-10^6$~K). This gas was likely  released in several episodes of SN explosions in massive star-forming regions. It was then ejected from the galaxy by stellar feedback in the form of filaments flowing out of the stellar disc~(see Fig.~\ref{fig::chains}), carrying metals into the surrounding gaseous halo which can then be mixed with the pre-existing gas. However, due to the low density of the gaseous halo, the dilution of metals is limited, and the resulting CGM abundances can be very large compared to the newly formed stars.  An important implication, is that the high-metallicity gas cannot take part in subsequent star formation for a considerable time. This means that there is a time delay between stars releasing metals and those metals returning to the ISM. On a longer time scale, this mixture of pristine and recycled material with large angular momentum is then accreted by the existing stellar component, contributing to the formation of a thin disc. 
In particular in models with a high early star formation rate~(models 1, 2, 3) stellar feedback is strong enough to remove low angular momentum gas from the disc and make it available in the halo where it is efficiently captured by the rotating gas, similar to the process noticed in a number of galaxy formation studies~\citep[see, e.g.,][]{2008MNRAS.387..577O,2010MNRAS.406.2325O,2014MNRAS.443.2092U,2020arXiv200303368K}. Then, after being mixed with surrounding CGM gas, the ejected gas falls back onto the disc, bringing more gas for further star formation during the thin disc formation epoch~\citep{2008MNRAS.386..935F,2009MNRAS.397.1804B, 2010MNRAS.404.1464M}.
To demonstate this, in Fig.~\ref{fig::enrich} we show the relation between the ISM metallicity and the spherical radial velocity, where negative velocities trace inflows while positive velocities trace outflow. At early stages of intense star formation the high metallicity regions are found preferentially in outflowing gas while infalling gas has lower abundances~\citep[see also][]{2020arXiv200907809P}. Thereby, when this gas is re-accreted onto the disc, it is already enriched, and not pristine, which appears to confirm various observations in the Milky Way~\citep[see, e.g.,][]{2014MNRAS.443.3809B,2015ApJ...800...14M,2016ApJ...816L..11F,2016ApJ...830...87S} and in external galaxies~\citep[see, e.g.,][]{2011Sci...334..948T,2011ApJ...733..111T}. This overall cycle is also very similar to the one described in \cite{2020arXiv200506310H}, who analysed EAGLE simulations of Milky Way-like galaxies, in particular for what concerns the early metal enrichment of the hot CGM by outflows, and the fact that most of the gas accretion during the thin disc phase is already recycled gas.

A notable feature in Fig.~\ref{fig::gas_temp}, is the two distinct areas of cold gas, that correspond to the formation of the thick disc, at high-$\alpha$ and to the thin disc, at low-$\alpha$. The high-$\alpha$ sequence is present as early as the first Gyr, as can be seen in the first row, while the low-$\alpha$ sequence appears later, at higher metallicity, indicating that the cold ISM has been gradually enriched in metals. Another notable feature is the absence of high-$\alpha$ abundances in the ISM after the thick disc formation (the two lower rows) which is likely caused by the mixing of metals in the CGM and the near galactic disc. 

Fig.~\ref{fig::cem} illustrates the range of chemical abundances of the ISM that exist at any given time in the simulation, due to the presence of gas at very different temperatures. It shows the chemical abundance evolution of newly formed stars (yellow curves) and the ISM (colour density maps) in model~1 in four 1~kpc-width circular annuli, while the other models are presented in the Appendix~\ref{app}~(see Fig.~\ref{fig::cem234}). As mentioned earlier, in all models the ISM abundances~($\FeH$ and $\aH$) can rapidly reach very high values, with a (moderate) fraction reaching solar metallicities already at $t<1$~Gyr. Meanwhile, stellar populations do not gain such a high metallicity before the firsts $\approx 2$~Gyr, even in the galactic centre. At large galactocentric radii, the difference between stellar and ISM abundances is even more evident, most of the ISM being more enriched than the population of locally formed stars. 

The large spread of ISM abundances at all radii clearly demonstrates that the assumption of instantaneous mixing~\citep{,2009A&A...504...87S,2014SAAS...37..145M} used in chemical evolution models is a crude approximation because it assumes that all metals are available and well mixed in the gas that forms stars at all times. From Fig.~\ref{fig::cem} it can be seen however that stars can form in an ISM which is substantially less (at large radii) or more (at small radii) metal-rich than the majority of the gas. Hence, in chemical evolution models, this will lead to chemical evolutions that are either faster or slower, depending on the location, than what the evolutions really is in more elaborate models.
In the inner and outer galaxy in particular, star formation occurs neither in the most enriched gas nor in the most widespread~(in terms of the gas mass abundances). Such behaviour is explained by the complex multi-phase structure of the ISM. Having illustrated the complex circulation of metals in the gas phase, we now turn to the emergence of the chemical patterns in stars.

\section{Chemical abundance patterns}\label{sec::results}

\begin{figure*}
\begin{center}
\includegraphics[width=0.6\hsize]{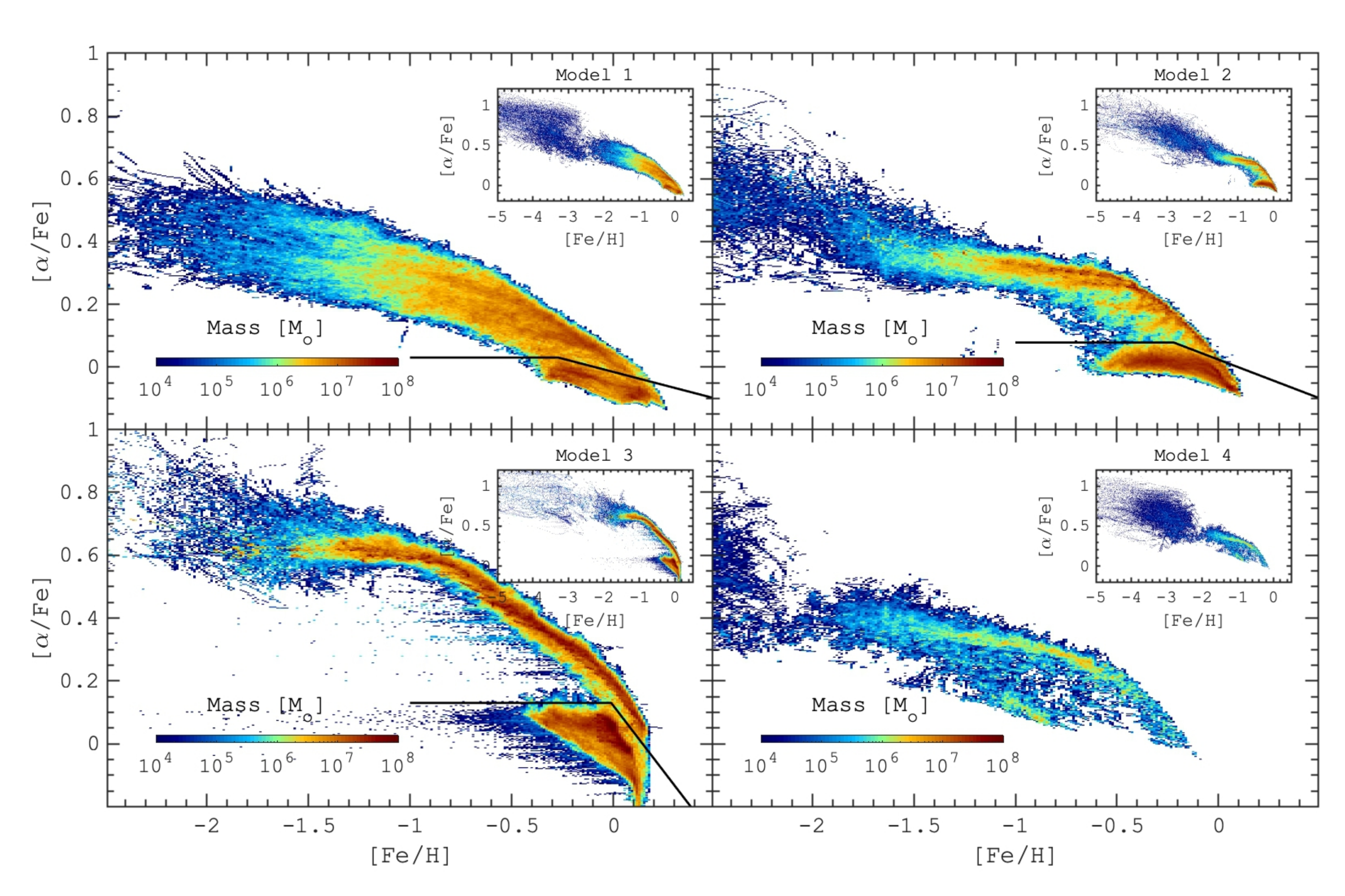}
\caption{$\aFe-\FeH$ maps of simulated galaxies at $10$~Gyr colour-coded by the stellar mass. Large frames correspond to a restricted chemical abundance range while smaller subframes depict the distribution on a larger metallicity scale to show where the oldest very metal-poor stars~($\FeH \lesssim-5...-4$) can be found. Black lines separate low- and high-$\alpha$ sequences in three models except model 4~(bottom right) where we do not find a prominent bimodality.}\label{fig::afe_all}
\end{center}
\end{figure*}

We begin by examining the global chemical abundance patterns in the simulated galaxies, in particular we focus on the $\aFe-\FeH$ distribution at the final snapshot $t=10$~Gyr. In Fig.~\ref{fig::afe_all} we show $\aFe-\FeH$ relation for all the stars in the four models. The relation for each model is shown in a restricted range of metallicities~(\FeH$>-2.5$ and, \aFe<1, large frames) and for a larger range of abundances~(\FeH$>-5$, \aFe<1.2) where the most metal-poor first generations of stars, being minor contributors to the total disc mass, are clearly visible. This figure shows that three~(models 1,2,3) among the four simulations produce clear low and high $\aFe$ sequences in the $\aFe-\FeH$ plane,  qualitatively similar to those observed in the Milky Way~\citep[see, e.g.,][]{2015ApJ...808..132H,2019arXiv191209778Q}. To further understand our results, we separate high- and low-$\alpha$ sequences~(see black lines in Fig.~\ref{fig::afe_all}) by a different lines for different models, except for model 4 which shows a single pattern similar to the high-$\alpha$ sequences of other models but shows no obvious low-$\alpha$ sequence. The low and high $\alpha$ sequences, as defined by this separating line, can be seen as tracing mostly the thin and thick discs, as is done in practice for observations of stars in the  Milky Way~\citep[see, also,][for simulated thin/thick discs analysis]{2012MNRAS.426..690B}.

\subsection{Low and high-$\aFe$ sequences: relation to structures (thin and thick discs)}

Discrimination between low- and high-$\alpha$ sequences allows us to connect the observed bimodality with structure and kinematics of these chemically defined stellar components of simulated galaxies. In particular, we look at spatial distributions of these populations and global differences in their kinematics. In Table~\ref{tab::tab} we summarize the main properties of simulated galaxies where we also compare some main parameters for stars belonging to low- and high-$\alpha$ sequences. Across different models the mass of the high-$\alpha$ population is rather similar $(3.4-3.8)\times10^{10}$~\Msun which correspond to $45-60$\% of the entire galaxy stellar mass. Note that these numbers are somewhat arbitrary and depend on the definition of the low- and high-$\alpha$ sequences. However, in all cases, our models show a significant high-$\alpha$ disc mass, which is in quantitative agreement with the estimate of~\cite{2015A&A...578A..87S} who found that 47\% of the stellar mass belongs to the high-$\alpha$ sequence in the Milky Way. In all three galaxies with bimodality, the high-$\alpha$ components exhibit shorter radial extend compared to low-$\alpha$ stars, and is also vertically thicker and dynamically hotter in both radial and vertical directions~(see Table~\ref{tab::tab}, note that the velocity dispersion components ($\rm \sigma_R, \sigma_z$) are measured for stars across the entire galaxy). Our simulations imply that, similarly to the Milky Way, the high-$\alpha$ sequence in models 1, 2 and 3 correspond to chemically defined compact thick discs, while the low-$\alpha$ stars form chemically defined extended thin discs~\citep[see, e.g.,][]{2012ApJ...753..148B,2015ApJ...808..132H,2019MNRAS.489.1742F}. It is also worth mentioning that, in our models, we find in-situ formed stars with chemical abundances typical of the low-$\alpha$ sequence discussed in the observational data and typically interpreted as made of accreted stars~\citep{2010A&A...511L..10N, 2018ApJ...863..113H,2018Natur.563...85H,2019A&A...632A...4D}. In other words, our simulations suggest that in-situ stellar populations of the Milky Way could contribute to the low $\alpha$-populations at metallicities below $\lesssim-0.5$.

\begin{figure}
\begin{center}
\includegraphics[width=1\hsize]{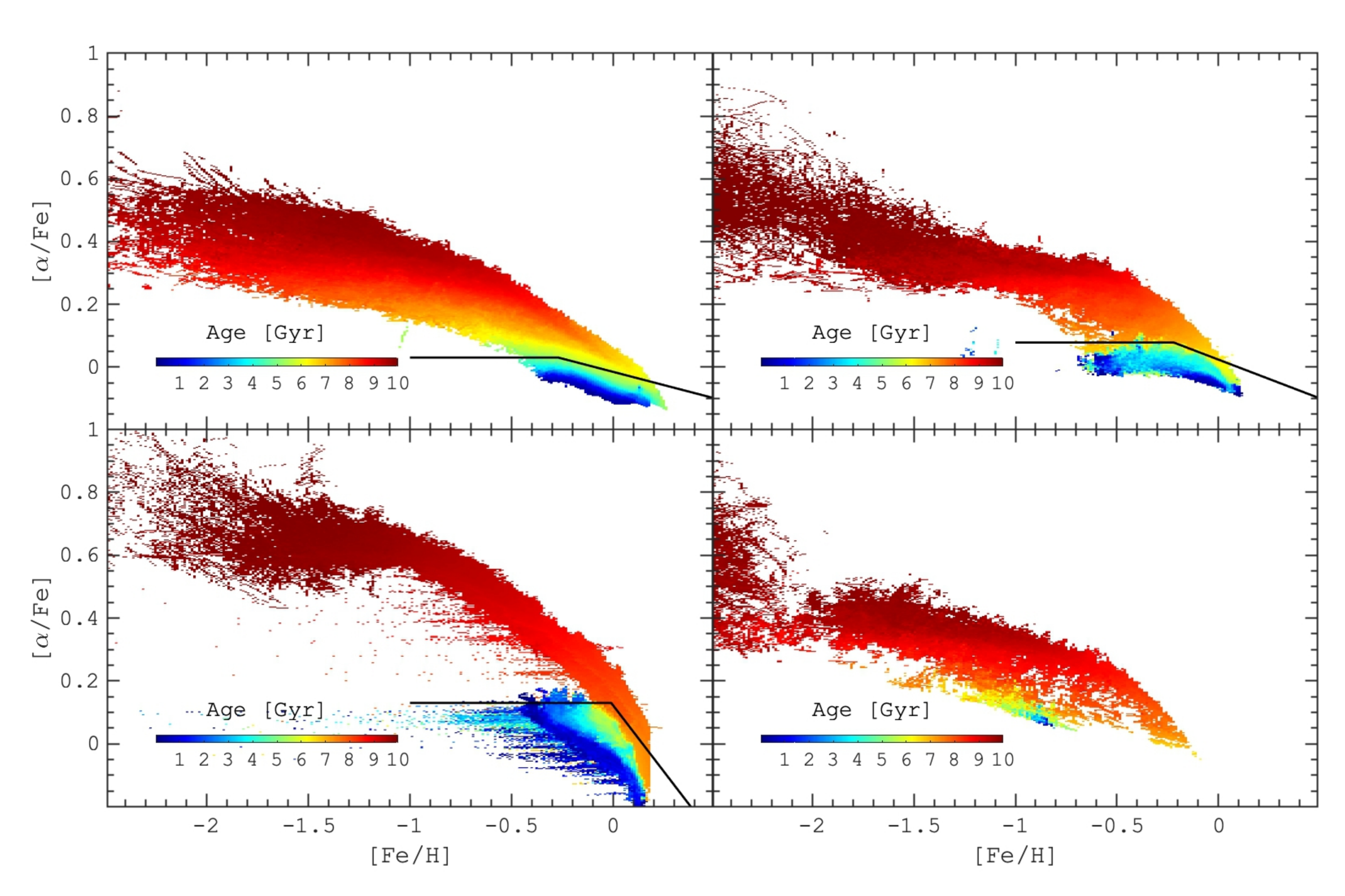}
\caption{Mean stellar age maps of $\aFe-\FeH$ relation in simulated galaxies at $10$~Gyr. Black lines separate low- and high-$\alpha$ sequences, as in Fig~\ref{fig::afe_all}. The distributions show that high-$\alpha$ sequence stars formed at early epochs of the galaxy evolution while low-$\alpha$ is the result of later evolution.}\label{fig::afe_age}
\end{center}
\end{figure}

\subsection{Low- and high-$\aFe$ sequences: relation to age}

The first question we want to address is related to the age differences between the two $\alpha$-sequences. In Fig.~\ref{fig::afe_age}, we show the $\aFe-\FeH$ distribution for all stars in the simulated galaxies colour-coded by the mean age of stars. The age structure of our galaxies is similar to the one observed in the solar vicinity~\citep{2013A&A...560A.109H}. In particular, all models~(1, 2, 3) with bimodality have a high-$\alpha$ sequence mainly formed during the first $2-3$~Gyrs of evolution, implying that the thick discs in our simulations are made of old stars. Moreover, it can be seen that the high-$\aFe$ sequence is clearly an age sequence, in the sense that age decreases along the sequence. This is particularly true for model 3, where the sequence is tight. This is also illustrated in Fig.~\ref{fig::tracks} where we present the chemical evolution tracks in the $\aFe-\FeH$ plane in $1$~kpc width circular annuli at different galactocentric distances built from recently formed stars~(age$<10$~Myr) stars as a function of time~\citep[see, also,][, Fig.8]{2012MNRAS.426..690B}.

Thin discs~(or low-$\alpha$ stars) contain stellar populations with a wider range of ages~($\lesssim 6-8$~Gyr) suggesting their formation on a longer time scale after the thick disc formation. The sharp age transition is in favour of a rapid change of the star formation regime leading to different enrichment efficiencies during the formation of high- and low-$\alpha$ sequences. A rapid transition between thick and thin discs formation phases also implies that in our simulations, both $\alpha$-sequences are not formed concurrently as it has been suggested in~\cite{2019MNRAS.484.3476C}. On the contrary, in our models, the thick and thin disc evolutions  are sequential~\citep[see also][]{2013MNRAS.436..625S}. In the outer disc however, where the thick disc does not exist, the star formation ignition will depend on the gas density, which will depend on a range of related processes, such as the accretion history at large radii, outflows from the inner disc, and/or the possible occurrence of a gas-rich merger.

Another feature related to the star formation history is the width of the high-$\alpha$ sequence. This sequence is 
tightest for the model with the most intense SFR during the thick disc phase~(model 3), due to the largest initial gas mass and smallest scale length. On the contrary, the model with the longest thick disc formation phase~(model 1) has a wide high-$\alpha$ sequence. Model 1 has radial stellar scale length for the thick disc which is 40-70\% larger than for the other models, or less concentrated gas distribution, implying less intense star formation during thick disc formation. The consequence is that the formation of the thick disc is much less concentrated than for the other models, less well mixed, with a clear radial dependence of the formation of the thick disc. Hence, the high-$\alpha$ sequence is composed of parallel chemical tracks.

\subsection{Low- and high-$\aFe$ sequences: relation to radius}
Next we look at the radial distribution of stars of both $\alpha$-sequences. In Fig.~\ref{fig::afe_rad} we show $\aFe-\FeH$ distributions but now colour-coded by the mean galactocentric distances for stars at the end of the simulation. Across all models, during the formation of high-$\alpha$ sequences, the chemical evolution is fast but slightly different at different radii due to the radial decrease of the star formation efficiency~\citep[see, e.g.,][]{2014ApJ...796...38N}. In particular, we observe parallel tracks reaching the same values of the \FeH delayed in time at a larger radius, stars forming more rapidly in the inner disc. This process is seen prominently in models 1, 2 and 4,  where star formation intensity follows the decreasing gas density (and hence increasing distance to the Galactic centre), leading to a chemical evolution which depends on radius. In model 3, the high star formation drives turbulent ISM conditions that contribute to a more uniform chemical abundance evolution across a range of radii. This leads to a tighter high-$\alpha$ sequence, but we still can distinguish parallel chemical evolution tracks~(see Fig.~\ref{fig::tracks}).

Meanwhile, thin discs show~(except for model 4) a negative radial metallicity gradient which is most prominent in models 1 and 2, the metallicity decreasing towards the outer disc as is apparent along the low-$\alpha$ sequence (see Fig. \ref{fig::afe_rad}). 
In these models, the formation of the disc is clearly inside-out, in the sense that the peak of star formation occurs at earlier times and is more intense in the inner disc. This is illustrated for model 1 in Fig.\ref{fig::cem} which shows the star formation history at different radii, and for model 2 in the Appendix. Hence thick discs~(high-$\alpha$) form initially in the inner galaxy, and thin stellar discs assemble later over a larger spatial scale. This is confirmed by Fig. \ref{fig::hayden1}, which shows the \FeH-\aFe distribution in different radial and vertical ranges. The figure clearly illustrates that the ridge line defining the high-$\alpha$ sequence is confined to the inner disc, while the full extent of the low-$\alpha$ sequence is visible only at larger radii, being restricted to high metallicities in the inner Galaxy. This behaviour is very similar to what is found in the Milky Way with the APOGEE data~\citep{2015ApJ...808..132H,2019arXiv191209778Q}.

\subsection{Fine structures}
Another interesting feature during the formation of the high-$\alpha$ sequence in model 2 is the horizontally-aligned ``spikes''. These brief, low-amplitude \FeH dilution episodes (see also in Fig.~\ref{fig::afe_all} as horizontal ridges) are likely associated with bursts of star formation during the thick disc formation phase~($<2$~Gyr, see Fig.~\ref{fig::sfr_infall} left). Note that similar patterns can be found when gas-rich mergers~\citep{2016MNRAS.456.3119S,2020MNRAS.491.5435B} stimulate the star formation in the low-\FeH gas accreted during the merger.

After the formation of the high-$\alpha$ sequences in models with bimodality~(1, 2, 3) we observe a rapid decrease and subsequent slow increase of the metallicity creating a ``loop'' similar to one predicted by the chemical evolution models by \cite{2017MNRAS.472.3637G,2019A&A...623A..60S,2020A&A...635A..58S}. In our simulations, the $\aFe-\FeH$ loop is the result of the star formation quenching being induced by the sharp change of the gas infall rate into the galaxy controlled by the stellar feedback~(winds). In Fig.~\ref{fig::sfr_infall}, we can see that the star formation quenching is the most rapid~($\approx 1$~Gyr) in model 3, being slower in model 1~($\approx 2$~Gyr) and model 2~($\approx 3$~Gyr). Thereby, the different quenching ``speed''~(and likely the star formation ratio before and after the quenching) is clearly imprinted in the diversity of the loop shapes across the models. In particular, in model 1, with a slow decrease of the star formation rate, the loop amplitude is small  and it results in a \FeH dilution of $\approx 0.1$~dex while in model 2 the loop is very evident, especially in the inner galaxy where the \FeH dilution is $\approx 0.2$~dex. Finally, in model 3, we found a \FeH dilution of $0.25-0.3$~dex due to the fastest transition between the intense star formation during the thick disc phase and quiescent thin disc formation phase. Note also that the amplitude of the \FeH dilution in our simulations is significantly smaller compared to $1.2$~dex in \cite{2017MNRAS.472.3637G} who assumed primordial chemical composition of the infalling gas. In our simulations the chemical composition of the infalling gas during the thin disc formation phase is not primordial because the gaseous halo has been significantly polluted during the thick disc formation, providing a tight connection between chemical abundance patterns in thick and thin discs. 

\begin{table*}
\caption{Initial and final~(at $10$ Gyr) parameters of simulated galaxies: gaseous halo mass~($\rm M_g(0)$), gaseous halo scale length~($\rm R_{gas}$), initial gaseous halo spin~($\lambda$); total stellar mass~($\rm M_{stars}$),  total number of star particles~($\rm N_{stars}$), $l$ is the radial disc scale, $h$ is the vertical disc scale, $\rm  \sigma_R$ and $\rm  \sigma_z$ are the mean radial and vertical velocity dispersions respectively. Upper indexes $(1),(2)$ correspond to low- and high-$\alpha$ sequences respectively relative to the cuts shown in Fig.~\ref{fig::afe_all}. Note that the velocity dispersion components are measured for stars across the entire galaxy and the values are not suitable for the direct comparison with the numbers in the Solar vicinity.}
\begin{tabular}{|*{20}{c|}}
\hline
\multicolumn{1}{|c|}{Model} & \multicolumn{3}{c|}{Initial parameters} & \multicolumn{8}{c|}{Final (at $10$ Gyr) parameters} \\ 
\hline
 	& $\rm M_{gas}(0)$ & $\rm R_{gas}$ & $\lambda$ & Morphology  & $\rm M_{stars}$  &  $\rm N_{stars}$ & $\Oo \rm M^{(1)}_{stars}/M^{(2)}_{stars}$ & $\rm l^{(1)}/l^{(2)}$ & $\rm h^{(1)}/h^{(2)}$ & $\rm \sigma^{(1)}_{R}/\sigma^{(2)}_{R}$ &  $\rm  \sigma^{(1)}_{z}/\sigma^{(2)}_{z}$
	\\
& $10^{10}$\Msun  & kpc	& &   & $10^{10}$\Msun   & $10^6$ & $10^{10}$\Msun & kpc & kpc & \kmps & \kmps
	 \\ \hline 
Model 1 & 6  & 3 & 0.0436 & barred &  5.8   & 4.6 &  2.33 / 3.46  & 7.3 / 5.5  & 0.29 / 0.48 & 54 / 67 & 32 / 34  \\ 
Model 2 & 9 & 5 &  0.0465 & barred &  8.2   & 4.7 &  4.77 / 3.42 & 4.3 / 3.2  & 0.35 / 0.59 & 73 / 87 & 27 / 37  \\ 
Model 3 & 9 & 3 & 0.049 & spirals & 8.4    & 4.1  & 4.57 / 3.82 &  5 / 3.9 & 0.22 / 0.38 & 69 / 104 &  30 / 49  \\ 
Model 4 & 6 & 5 & 0.0347 & spirals &  3.4    & 2.4 & - / 3.4 & - / 3.3 & - / 0.24 & - / 51 & - / 27 \\ \hline
\end{tabular}\label{tab::tab}
\end{table*}

\subsection{Origin of the bimodal sequence}

Although all models present a variety of chemical patterns, the first three have conspicuous bimodal sequences, and it is possibly embryonic in model 4 (weakly visible on Fig. \ref{fig::maps_at_end}). The high-$\alpha$ sequence is a temporal sequence because metallicity increases and alpha abundance decreases monotonically with time during this phase (see Fig. \ref{fig::afe_age} and Fig. \ref{fig::tracks}). The low-$\aFe$ sequence is somewhat more complex and result from the combination of two different effects: the first is the accumulation of stars younger than 6-7 Gyr at the end of each chemical track. This is seen in Fig. \ref{fig::tracks}, which shows the chemical evolution in different parts of the simulated galaxies. We measure the chemical evolution tracks as the evolution of the chemical abundances~($\FeH$, $\aFe$) for young stars~(age<10~Myr) in 1~kpc-width circular annuli at four galactocentric distances $0, 4, 8, 12$~kpc. Because metallicity and $\alpha$-abundance evolve little during the thin disc phase, reaching equilibrium values, thin disc stars pile up at the end of each track. The second effect contributing to the formation of the low-$\alpha$ sequence is the gradual shift of these equilibrium metallicities to lower values at larger radii. This gradual shift builds up a continuous sequence from low to high metallicity.

\begin{figure}
\begin{center}
\includegraphics[width=1\hsize]{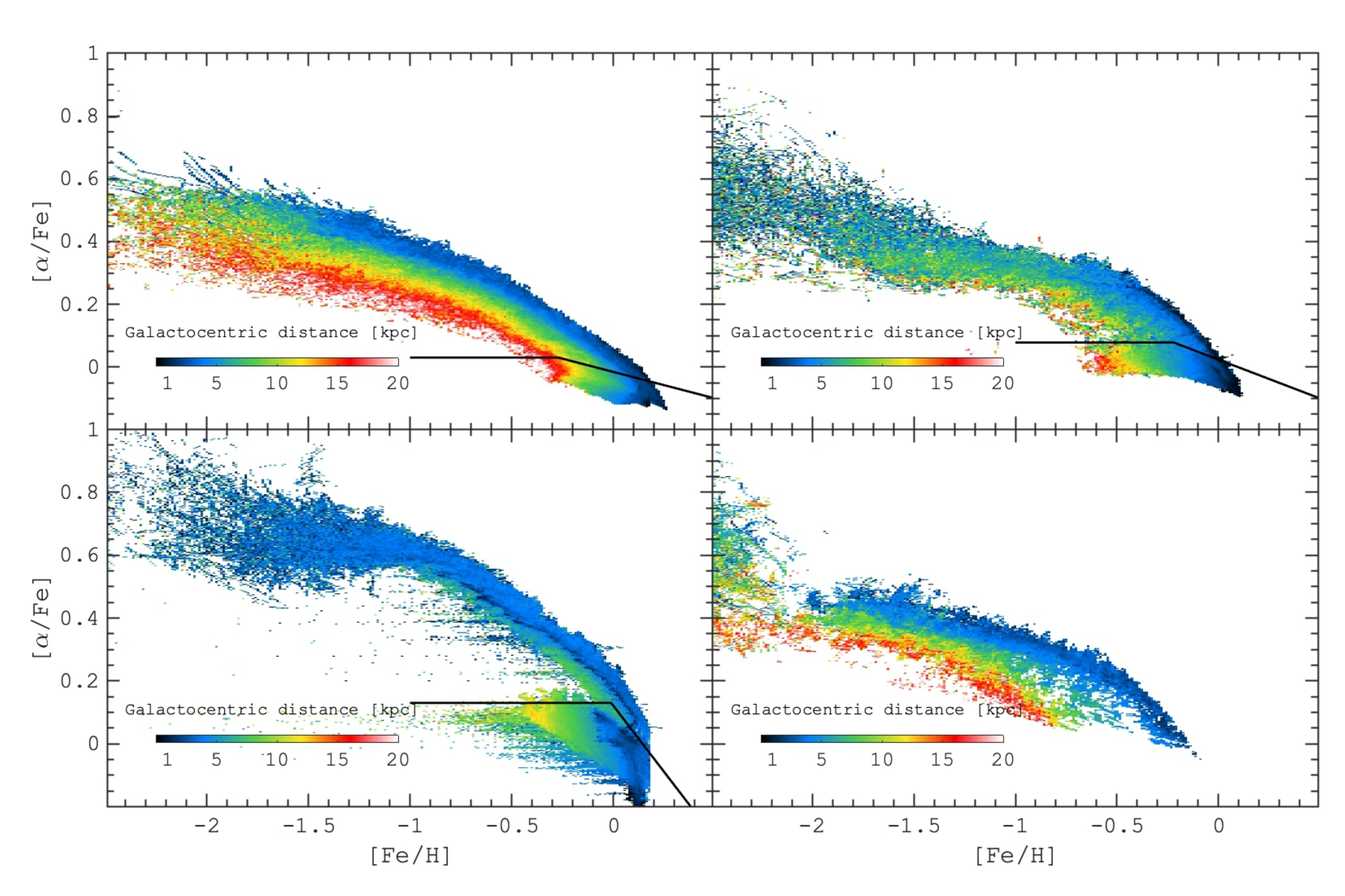}
\caption{Mean galactocentric distance maps in the $\aFe-\FeH$ plane in simulated galaxies at $10$~Gyr. Black lines separate low- and high-$\alpha$ sequences, as in Fig~\ref{fig::afe_all}. Radial metallicity gradient is visible across all the models on the low-$\alpha$ sequence, but also on the high-$\alpha$ sequence for model 1, 4, and to a lesser extent, on model 2. It is essentially absent in model 3.}\label{fig::afe_rad}
\end{center}
\end{figure}

\begin{figure*}
\begin{center}
\includegraphics[width=0.8\hsize]{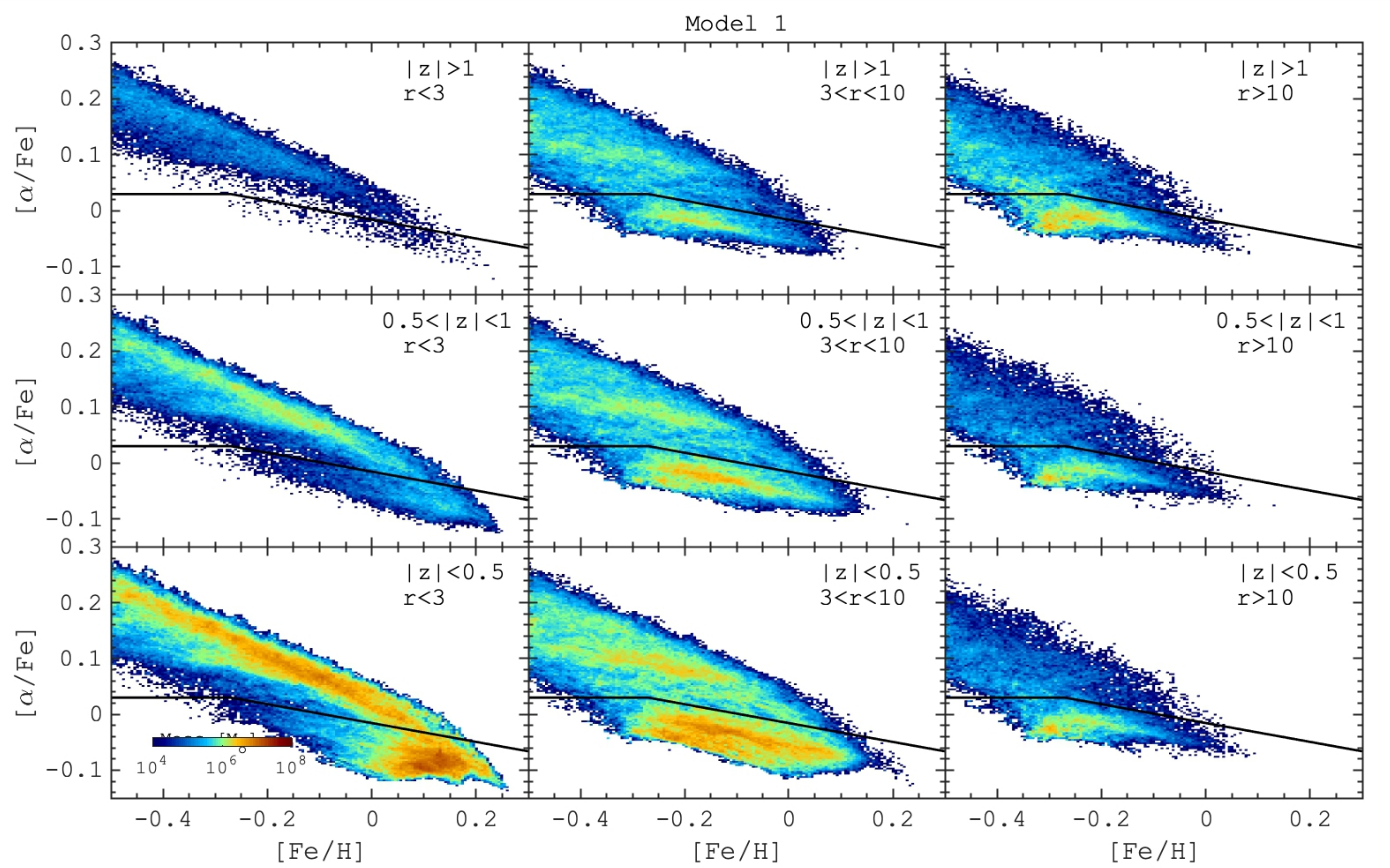}
\caption{Spatial variations of the $\aFe-\FeH$ maps in different parts of the disc at the end of the simulation in model 1. 
The low-$\alpha$ sequence disappears far away from the galactic plane, except in the outer disc due to flaring~\citep[see][]{2014A&A...572A..92M}, while the high-$\alpha$ sequence becomes barely visible at large galactic radii. This $\aFe-\FeH$ variations show that similar to the Milky Way~\citep[see, e.g.,][]{2015ApJ...808..132H,2019arXiv191209778Q}, the high-$\alpha$ sequence corresponds to radially compact thick disc, while the low-$\alpha$ stars represent the radially extended thin disc component. We note that although the SFH shows no discontinuity, the \aFe-\FeH shows clearly two distinct sequences dominated by the thick and thin discs. Spatial variations of the $\aFe-\FeH$ relation in other models are shown in Appendix~\ref{app}~(see Fig.~\ref{fig::hayden234}).}\label{fig::hayden1}
\end{center}
\end{figure*}

\begin{figure*}
\begin{center}
\includegraphics[width=0.8\hsize]{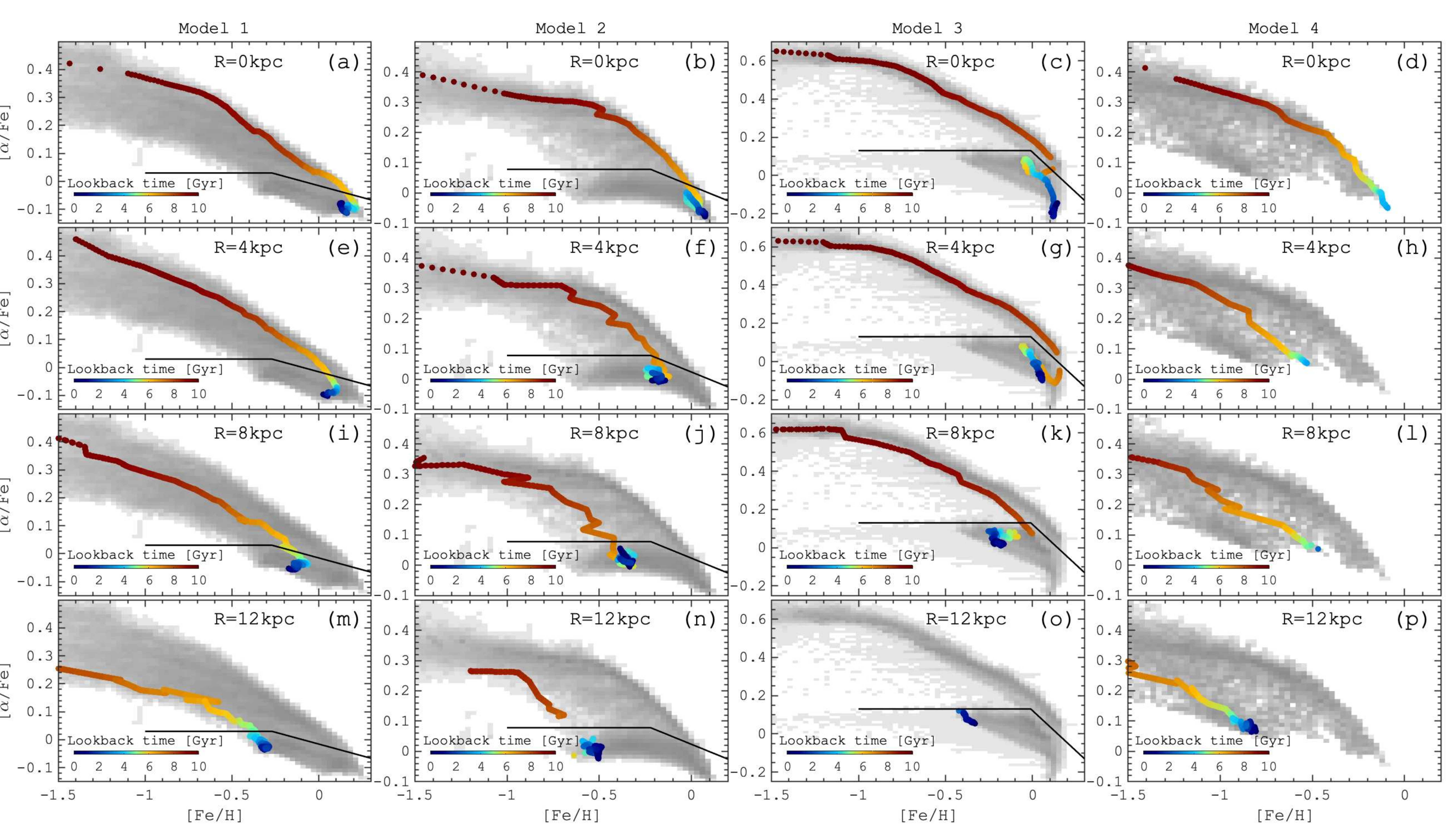}
\caption{Chemical evolution tracks in $\aFe-\FeH$ plane in $1$~kpc width circular annuli at different galactocentric distances built from newly born~(age$<10$~Myr) stars as a function of time. Grey background depicts the entire $\aFe-\FeH$ relation for all the stars at the end of the simulation. Each galaxy formation model is shown in a separate column. Models with a clear bimodality~(1, 2 and 3) have endured a rapid (and moderate) dilution of \FeH during the decrease of the star formation rate~(see Fig.~\ref{fig::sfr_infall}). This episode is imprinted as a non-monotonic stretch of the chemical evolution track, or even a loop~\citep[see also][]{2017MNRAS.472.3637G,2020A&A...635A..58S} being  the most prominent in models 2 and 3.}\label{fig::tracks}
\end{center}
\end{figure*}

The existence of the low and high-$\alpha$ sequences is a natural consequence of stars forming in two different ISM: one in which chemical species are produced in an intense episode of star formation and well mixed, which can occur only in primordial, concentrated discs, and the other in more extended discs, where the star formation efficiency is lower and follows the gas density, decreasing towards the outer disc. It is likely that these two types of ISM in galaxies are determined primarily by the presence of a substantially massive thick disc since the low-$\alpha$ sequence, and its ridge line should occur whenever the gas density decreases radially, which is expected in more or less all galaxies with a thin disc. Moreover, the bimodality goes with the dichotomy in radius, with the high-$\alpha$ dominating in the inner disc and the low-$\alpha$ at larger radii. When the infall rate provides only a limited amount of gas, as in model 4, it leads to the formation of a massive thick disc and no clear thin disc component. In which case the dichotomy in the $\aFe-\FeH$ plane is absent, or barely visible.

\begin{figure*}
\begin{center}
\includegraphics[width=0.8\hsize]{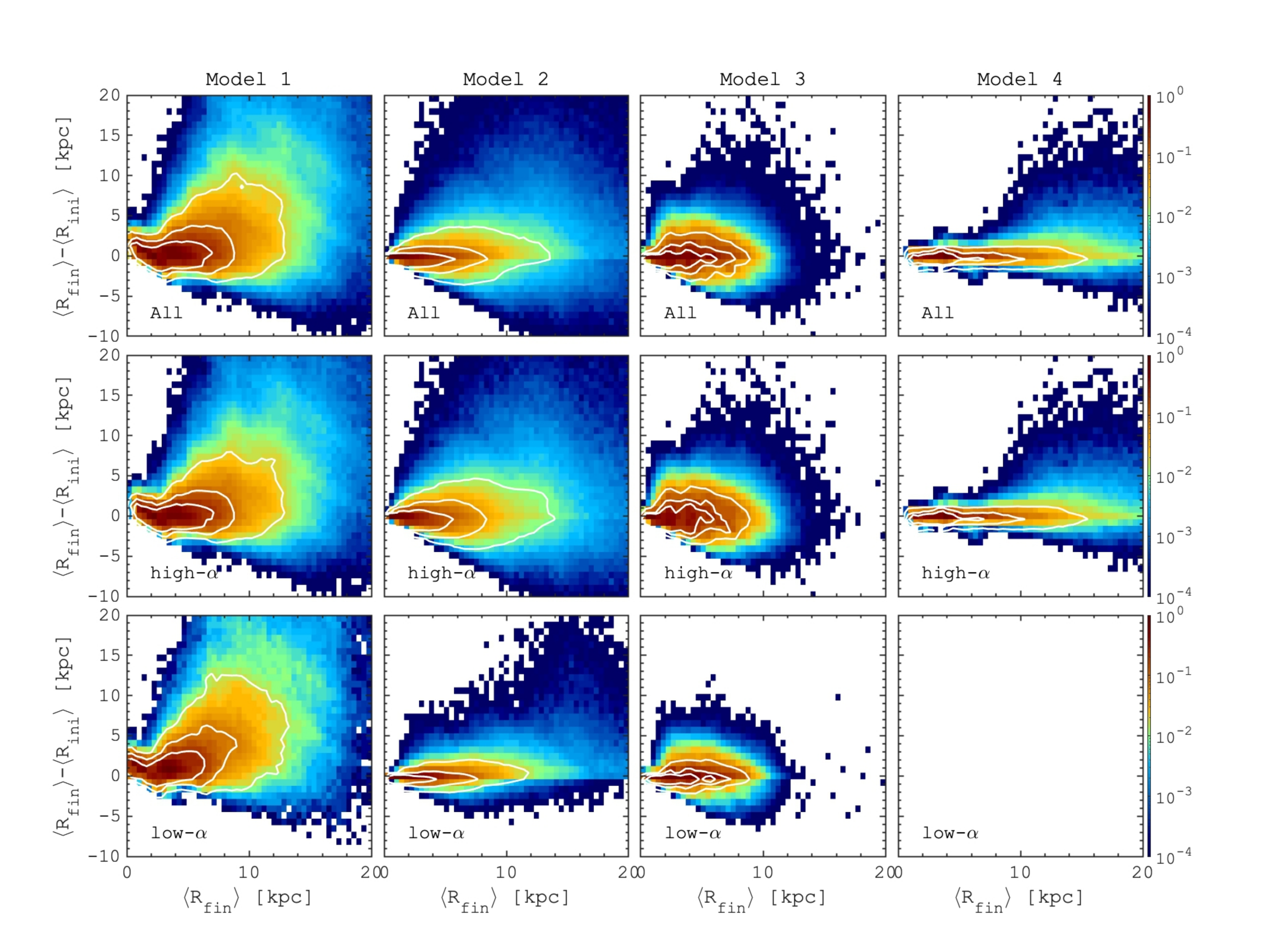}
\caption{Distributions of the mean change of the galactocentric distance $\Rini-\Rfin$ as a function of \Rini for different models: top row corresponds to all stars while the middle and the bottom to high- and low-$\alpha$ sequences respectively. Initial galactocentric position of star particles~(\Rini) is measured as the mean position during the first $500$~Myr after a star particle formation. Final galactocentric position of star particles~(\Rfin) is measured as the mean position during the last $500$~Myr before the end of simulation~($10$~Gyr). Maps are shown in the same scale and colour-coded according to the mass fraction relative the maximum value. Three white contours in each frame correspond to 90\%, 70\% and 50\% of the enclosed mass. Across models, stellar radial migration is limited to $4-5$~kpc for $90\%$ of the stars suggesting that only $10\%$ of stars are extreme migrators.}\label{fig::migration1}
\end{center}
\end{figure*}

All four models illustrate that significant differences can arise in the chemical patterns because of the different star formation histories, and spatial distribution of the star formation occuring within the modeled galaxies. The star formation histories are themselves different because of the initial distribution of the gas and its total mass in each galaxy (see table\ref{tab::tab}). These initial conditions result in a factor of 5 difference in gas surface densities, explaining the different star formation intensities reached in the first Gyrs of evolution by each galaxies. It clearly illustrates how differences of the order of 30\% in the initial gas mass and its spatial extent produce significant variations in the final chemical patterns.

\section{Role of radial migration}\label{sec::migration}

Although it is widely accepted that radial migration plays a role in reshaping the structure of galaxies over time
~\citep[see, e.g.,][]{2008ApJ...675L..65R,2009MNRAS.396..203S,2011ApJ...737....8L,2012A&A...548A.126M,2012MNRAS.426.2089R,2013MNRAS.436.1479K,2013A&A...553A.102D,2019ApJ...882..111D}, some studies suggest that, at least in the Milky Way, the global characteristics of the stellar disc do not seem to require significant migration, if only to explain the few percent of metal-rich and metal-poor stars on circular orbits at the solar vicinity~\citep{2013A&A...560A.109H,2015A&A...578A..58H,2018A&A...616A..86H,2019A&A...625A.105H,2020A&A...638A.144K}. Very limited radial migration seems also to be found in models where the coupled effects of bar and spiral arms is taken into account~\citep[see][their Fig 11]{2016MNRAS.461.3835M}. Also recent works based on the analysis of GES data seem to suggest that migration is limited, with for example, only 2\% of stars in the solar vicinity being bona fide candidate migrators from the inner disc, these stars having both high metallicity~($\FeH>0.1$) and circular orbits~\citep{2018A&A...609A..79H}.

\begin{figure*}
\begin{center}
\includegraphics[,width=0.7\hsize]{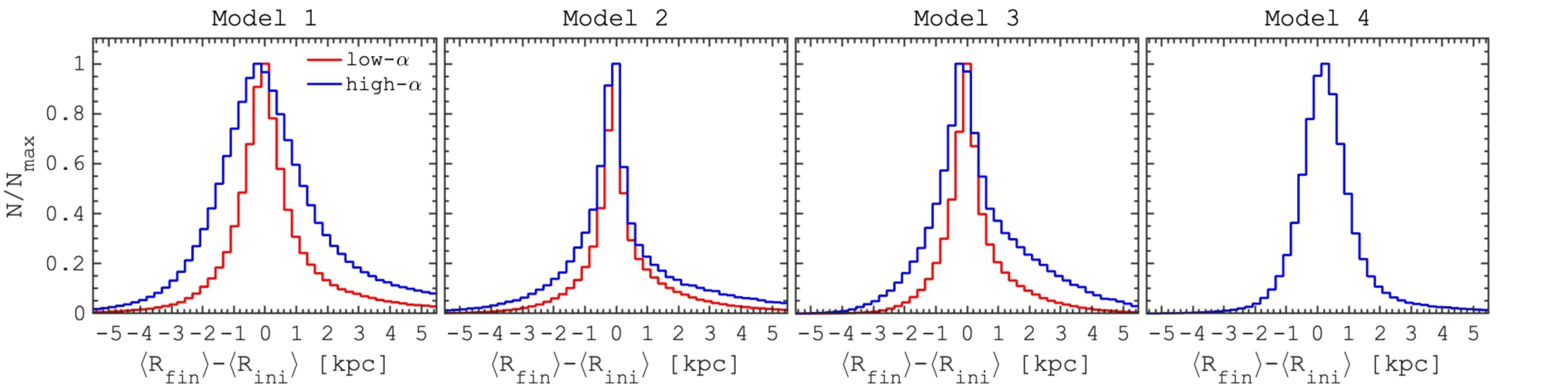}
\caption{Distribution of the difference between the mean final~($\rm \langle R_{fin} \rangle$) and the mean initial~($\rm \langle R_{ini} \rangle$) galactocentric distance for high-$\alpha$~(bl,ue) and low-$\alpha$~(red) stars in simulated galaxies. 
Difference between the mean final and initial positions of stars depicts the amplitude of the radial migration~(churning). Both thick and thin disc stars experience churning of $2-4$~kpc, but for a greater fraction of stars in the thick disc~(high-$\alpha$ sequence), which likely suggests a higher radial migration efficiency during the early phases of the galaxy evolution. The skewness of the distributions reflects the dominance of the outward churning caused by the negative density gradient in stellar discs.}\label{fig::migration2}
\end{center}
\end{figure*}

Our simulated galaxy sample includes galaxies with different morphologies~(see Fig.~\ref{fig::maps_at_end}), thus providing us with an opportunity to test how much radial migration is experienced by stars due to the different asymmetries and how they affect chemical patterns. In this section we first take a brief look on the efficiency of radial migration for thin~(low-$\alpha$) and thick~(high-$\alpha$) disc stars and we study the impact of radial migration on the chemical abundance patterns, and in particular on the $\aFe-\FeH$ relation over time. 

\subsection{The intensity of radial migration in our simulations}

First, we measure the amplitude of the radial displacement of stars in different models. In Fig.~\ref{fig::migration1} we show the relation between the change of the galactocentric distance~($\Rfin-\Rini$) as a function of the initial galactocentric distance~(\Rini). For this and the following figures, we define \Rini as the mean galactocentric radius of stellar particles,  averaged over a time interval of $500$~Myr~(50 snapshots) since their formation; in a similar way, \Rfin is the mean galactocentric radius of stellar particles, averaged over the last $500$~Myr, before the end of simulation. This approach allows us to estimate the churning of stars in kpc independent of the angular momentum calculation which can also experience periodic variations with zero net change, especially in the non-axisymmetric potential of the barred galaxies~\citep[see, e.g.,][]{2007MNRAS.379.1155C}. In Fig.~\ref{fig::migration1} we see that half of the stars migrate only up to $\approx2$~kpc from their birthplaces, about 40\% can be churned up to $\approx 4-5$~kpc while the most extreme migrators~($|\Rfin-\Rini|>5$~kpc) represent only $\approx 10\%$ of the total disc mass~\citep[see a similar behaviour in][]{2013A&A...558A...9M}. A notable feature is that radial migration is substantially more efficient in models with bars~(model 1 and 2) where stars can migrate by more than $5$~kpc. Model 3, without a prominent spiral structure, shows the less efficient overall churning, while in model 4, with spiral arms, migration is more efficient in the outer disc~($>8-10$~kpc) where the resonance of slowly rotating spirals is located. In our simulations, bar formation causes considerable more angular momentum changes, especially in the inner disc, than those of spiral patterns.

In Fig.~\ref{fig::migration2} we show how much radial migration is experienced by stars of both high- and low-$\alpha$ sequences or, as we showed above, chemically defined thick and thin stellar discs, respectively. It shows that the galactocentric distance change is dominated by thick disc stars (in particular in model 1 and 3), although the tails of the most substantial changes of guiding radii are almost equally extended in each population. This is in qualitative agreement with studies by \cite{2012MNRAS.422.1363S} and \cite{2018A&A...616A..86H}, who also found that churning in a thick disc is only mildly more important than the churning of thin disc stars. 
This figure shows that churning effects in the thin disc (as defined by stars belonging to the low-$\alpha$ sequence), is minimal, and affects a small number of stars. Finally, although we observe both inward and outward radial migration for both thick and thin disc stars, outward migration is slightly more significant in all our models due to negative radial density gradient~(see also Fig.~\ref{fig::migration1}). 

\subsection{Effect of radial migration on the chemical patterns}

We now test how much chemical abundance patterns in the \aFe-\FeH plane and metallicity distributions are affected by radial migration. Figure ~\ref{fig::migration3}~(top two rows) shows $\aFe-\FeH$ distributions for stars at their time of formation~(top row) and at the end of the simulations~(middle row) in the middle disc region~($5<R<10$~kpc) near the galactic plane~($|Z|<0.5$~kpc). Generally speaking in Fig.~\ref{fig::migration3}, radial migration only slightly smooths the $\aFe-\FeH$ distribution -- mostly by bringing some stars from the inner and outer disc regions -- but the distributions are only very marginally affected by migration, in all simulations, and whatever the structure dominating the asymmetries (bar or spiral arms). 

The typical radial excursion due to blurring (epicyclic oscillation) is of the order of $2$~kpc for an old high-$\alpha$ star. If radial mixing is due essentially to blurring, then chemical patterns sampled on a $2$~kpc radius range will be unchanged. If churning has a significant effect, chemical patterns will be substantially modified due to contamination by stars churned to a certain disc region. Fig.~\ref{fig::migration4} shows the same plots as the previous figure, sampled on a smaller radius range, going from $7.5$ to $8$~kpc. The comparison of the top and middle rows illustrate that churning has no significant impact. Similar to the Solar vicinity region in the Milky Way however, Fig.~\ref{fig::migration3} or  Fig.~\ref{fig::migration4} show that the most evident manifestation of the outward radial migration, visible in all models, is the presence of high-metallicity stars, likely formed in the innermost regions~\citep[see, e.g.,][]{1972ade..coll...55G,2015ApJ...808..132H,2015MNRAS.447.3526K,2018A&A...609A..79H}. To highlight this effect the bottom rows of the two figures show the metallicity distributions of stars at their initial~(dashed lines) and final radius~(solid lines) for high- and low-$\alpha$ sequences separately. Apart from this small shift in metallicity, the impact of radial migration is barely visible when comparing $\FeH$ distributions of stars where they are formed and at the end of the simulation. The contribution of these metal-rich stars does not exceed $1-2$\% of the total mass of both disc components, it is similar to the findings of other recent studies~\citep{2019A&A...625A.105H,2020A&A...638A.144K}.
 
These figures show why, in galaxies with a tight high-$\alpha$ sequence, such as in model 3, but also as in the Milky Way, the effect of radial migration on chemical patterns is limited. The high-$\alpha$ sequence remains unchanged because of the formation of the thick disc is mostly independent of radius. Hence, even though stars can migrate on substantial distances after formation, the high-$\alpha$ chemical patterns will not be affected, since they are similar at all radii, because of the efficient mixing occurring at early times. On the contrary, the low-$\alpha$ sequence depends on the radius of formation of the stars, but it is left unaffected by radial mixing because churning is limited in the thin disc.

\section{Discussion}\label{sec::discuss}

In this section, we first start by summarizing our results, then discuss the consequences of our findings.

\subsection{Our results}

The existence of a high-alpha sequence with a clearly visible ridge line (Fig. \ref{fig::afe_all}) is related to the presence of a dense and massive disc of gas. Models 2, 3 and 4, which all have this feature, have their high-alpha sequence associated with a disc of a scale length of 3.2 to 3.9 kpc, significantly smaller than the scale length of model 1, which is 5.5 kpc, nearly two times larger. The scatter of high-$\alpha$ sequence is more prominent in model 1, where the initial gas distribution is radially extended, thus leading to a progressive decline of the star formation efficiency with radius~\citep[see also][]{2014ApJ...796...38N} within the thick disc. Although, our models do not include the evolution of galaxies in cosmological context, our chemical abundance trends are in general agreement with 
several studies that include cosmological conditions with early mergers~\citep{2013A&A...554A..47G,2013MNRAS.436..625S,2016A&A...587A..10M}. This similarity suggests that mergers are not required for the thin/thick discs $\alpha$-bimodality while being essential for the chemical trajectories of accreted and in-situ halo stars of the Milky Way~\citep{2010A&A...511L..10N,2018ApJ...852...49H,2018ApJ...863..113H,2018Natur.563...85H,2019NatAs...3..932G,2020MNRAS.495.2645B}.

The evolution of the low-$\alpha$ sequence follows two possible paths. Firstly, the low-$\alpha$ sequence is made of the ends of the chemical tracks that start at high-$\alpha$ and low metallicity and go to low-$\alpha$ and high metallicities (see Fig. \ref{fig::afe_rad}).
In the last $7-9$~Gyr, change in metallicity and $\alpha$ is limited, reaching an ``equilibrium'' value, allowing stars to pile up at more or less the same position in the \FeH-\aFe plane. A negative gradient of the gas density drives a slower evolution towards the outer disc, the ends of the chemical tracks at larger R reaching lower values of metallicities. As a consequence of these two effects, the terminal metallicity at which the stars pile up gradually moves to lower values towards the outer disc, building the low-alpha sequence.
Secondly, the low-$\alpha$ sequence may also result from a loop in the \FeH-\aFe plane, when the disc ISM metallicity decreases at the end of the thick disc phase. This dilution results from the abrupt quenching of the star formation activity, as best illustrated by model 3. The quenching induces a drop in the amount of metals ejected to the halo. Then this gas rains back onto the disc with a lower metallicity, inducing a dilution. From Fig. \ref{fig::cem234}, it can be seen that the extension of the loop (and therefore of the low-$\alpha$ sequence) to low metallicity seems to be a direct function of the speed of the drop in the star formation activity. The drop in metallicity is about $-0.5$~dex when the drop in the star formation rate is almost instantaneous, as is the case in the radius range centred on $R=8$~kpc. 

When the formation of the low-$\alpha$ sequence results from the first path, the presence of these two sequences in a single galaxy is due to two difference modes of star formation, mainly driven by the gas density. The formation of a tight high-$\alpha$ sequence, is due to the high star formation intensity reached in massive discs, and depends on the early accretion of a large quantity of gas. The formation of a low-$\alpha$ sequence (as described by the first process) is expected to occur as soon as star formation proceeds at lower gas densities, outside the compact gaseous disc that formed the thick disc. The formation of a radially extended disc is expected in most cases independently of the particular merger history of the galaxy. Stars will then form at a rate which depends on the radius, and the mixing (mostly blurring) of stars in the disc will then generate a low-$\alpha$ sequence by blending together stars that are formed at different radii (hence at different metallicities).

Although we emphasize that nothing prevents the two sequences to be formed contemporaneously, stars will form first where the cold ISM reaches the necessary density, which occurs faster in the inner disc. This is visible looking at the star formation histories at different radii in the first 3 models (see Fig. \ref{fig::cem234}). In these models, the formation of the outer disc is delayed by a couple of Gyr. 
On the other hand, we do not confirm the parallel formation of both $\alpha$-sequences demonstrated in \cite{2019MNRAS.484.3476C}, which could possibly occur in our simulations if the multi-phase structure of the ISM was not taken into account. In particular, allowing star formation in a warm ISM~($T\sim10^4$), we would obtain a similar result~(see Fig.~\ref{fig::gas_temp} where the low-$\alpha$ sequence appears to be filled by enriched hot gas at early times). However, our results suggest that chemical evolution models should include multi-phase ISM evolution to correctly take into account the delayed metal enrichment of the star-forming gas~\citep[see, e.g.,][and Snaith et al. in prep]{2017A&A...605A..59R}.

What our simulations demonstrate is that the bimodality is a natural outcome in massive galactic discs. If early gas accretion has been sufficiently important in central regions of discs, and we argue below that this is probably the case for most Milky Way-like galaxies (guaranteeing the formation of a high-$\alpha$ sequence), then bimodality is assured, because low-gas density is always present in the outskirts of exponential discs. We emphasize that bimodality is conspicuous even when the global SFH of a galaxy shows no obvious separation, or distinct phases~(see for example model 2). This is because the two regimes of high and low star formation intensity are spatially separated, the high-$\alpha$ building up in the inner disc, while the low-$\alpha$ builds up mostly outside the inner disc.

\subsection{How do these results help understanding the evolution of the Milky Way?}

How do the simulations presented here help us to understand the evolution of the Milky Way? The Milky Way differs from all models in various ways, but some of its characteristics can be found in each model. For example, the tight high-$\alpha$ sequence observed in model 3, with weak deviations of the chemical evolution during high-$\alpha$ sequence formation observed at different radii, is very similar to the one we have in our Galaxy. In the MW, the small dispersion in $\aFe$ at a given age is in favour of a thick disc formation in a well-mixed turbulent ISM~\citep{2013A&A...560A.109H,2015A&A...579A...5H,2019MNRAS.489.1742F}. This is also consistent with the fact that galaxies at high redshifts are known to be gas-rich, and to have high gas mass surface densities~\citep{2010ApJ...713..686D,2010ApJ...718..177A,2015ApJ...800...20G,2016NatAs...1E...3P} and intense star formation~\citep{2009ApJ...699.1660L,2013A&A...555A..72L,2011ApJ...742...11S}.

On the other hand, the low-$\alpha$ sequence of the MW looks more like those seen in models 1 and 2, which in both cases are built through the first path described above: they are a composite of chemical tracks resulting from the evolution at each radius. This is the scenario described in \cite{2019A&A...625A.105H} and confirmed by the simulations presented here. \cite{2019A&A...625A.105H} found that while the high-$\alpha$ sequence is a temporal sequence, in the sense that age varies monotonically with metallicity and $\alpha$-abundance~\citep[see also][]{2013A&A...560A.109H}, the low-$\alpha$ sequence is a spatial sequence, in the sense that, at a given age, metallicity varies with radius along the ridge line of the sequence.  This corresponds to the way the low-$\alpha$ sequence is built in model 1 and 2: it is a composite of evolutions that proceeded almost independently at each radius. Then radial wandering of stars effectively built a sequence visible where this mixing is sufficiently effective (for example at the solar radius). The formation of the two distinct sequences may be strengthened by internal dynamical effects. Hence, in \cite{2015A&A...578A..58H}, we showed that the outer Lindblad resonance of the bar creates a separation between the inner and outer discs, thereby diminishing gas and star exchanges, fostering different evolution of the inner and outer discs. In simulations, stars form even at very low metallicities and very old ages in the outer disc, as illustrated by the chemical tracks of young stars shown in Fig. \ref{fig::tracks},  and provide the metals for the younger generations that dominate the low-$\alpha$ sequence, at ages less than $7-8$~Gyr. In the case of our Galaxy, there is yet no compelling evidence as to from where the oldest stars in the outer disc obtained their chemical abundances. Apart from the gas outflows expected during the thick disc formation \citep{2014ApJ...789L..30L} there is also ongoing, rapid cold gas accretion suggested by cosmological simulations~\citep{2003MNRAS.345..349B,2006MNRAS.368....2D,2009MNRAS.397L..64A,2014MNRAS.442..732W,2018A&A...610A..75C}. Another possible source is the gas deposited by early mergers (Sausage-Gaia-Enceladus~\citep{2018Natur.563...85H,2018ApJ...863..113H,2018MNRAS.478..611B}, Sequoia~\citep{2019MNRAS.488.1235M}, Kraken~\citep{2019MNRAS.486.3180K}). However, ~\cite{2019A&A...632A...4D, 2020MNRAS.494.3880B} found that the GES merger contributed to the heating of a fraction of the thick disc, which started to form before and continued its formation sometime after the last merger. Of course, some amount of gas could have been delivered by the previous mergers~(Kraken, Sequoia) however,~\cite{2019MNRAS.486.3180K,2020MNRAS.498.2472K} estimated that these satellite galaxies were even less massive. Therefore, the MW-type galaxies had the same~(still high) gas fraction, the amount of gas delivered by mergers should be an order of magnitude lower compared to the main progenitor. 

\subsection{Comparison with other recent studies}
There is another example of chemodynamical simulations that clearly show two sequences in the distribution of stars in the \FeH-\aFe space (oxygen in this case) are the MAGICC simulations in \cite{2016MNRAS.456.3119S}. In this study, the authors study the effect of various feedback prescriptions on the evolution of disc galaxies and in particular on the chemical patterns, starting the evolution with the same initial conditions. The galaxy evolved with the strongest early feedback prescription shows chemical patterns that look very similar to what we have in the Milky Way, with a high-$\alpha$ sequence that dominate their inner simulated galaxy, and a low-$\alpha$ sequence which is dominated by the thin disc at ages younger than 6-7 Gyr. Finally, the low-$\alpha$ is clearly radially dependent, as illustrated in their Fig.~6. Similar to our findings~\cite{2019MNRAS.486.3180K,2020MNRAS.498.2472K} show that smooth accretion supplies most of the gas for forming thin discs, while thick discs emerge from the much more turbulent gaseous phase at high redshift, fueled by gas-dominated mergers in their case.

More recently, \cite{2018MNRAS.474.3629G} explored the Auriga cosmological zoom simulations for a number of Milky Way mass galaxies, but the bimodalities that are described by these authors in the inner or outer discs bear more resemblance to the second type of bimodality described in the introduction. That is, the bimodality is generated by two distinct phases of star formation, the gap in the star formation activity in the outer disc being induced and maintained by a contraction of the gaseous disc. The resulting general patterns are much less evocative of a high- and low-$\alpha$ sequence as can be seen, for example, in  \cite{2016MNRAS.456.3119S}. In particular, the low-$\alpha$ sequence is not a ridge line that builds up progressively inside-out as the result of the accumulation of stars at an equilibrium metallicity, as is observed in  \cite{2016MNRAS.456.3119S}, or in this study. 

In our simulations, we showed that the low-$\alpha$ sequence forms when star formation efficiency ceases to be dominated by the formation of the thick disc, specifically in the radially extended thin disc. It de facto produces separate evolutions, even though the outflows generated by the thick disc may contribute to the metal enrichment of the outer disc. It is also the conclusion reached by \cite{2018MNRAS.477.5072M} that the pathways producing the two sequences in galaxies from EAGLE simulations are distinct. The formation of the two sequences, however, differs in several aspects from our work. For example in our simulations, and in the MW, the high-$\alpha$ sequence can be subdivided into two segments: the first corresponds to the formation of the thick disc, the second to the formation of the inner thin disc at high metallicities. In \cite{2018MNRAS.477.5072M}, this second segment does not exist, and the evolution in the inner disc seems to halt after a few Gyr, and explains the relatively small extension of the high-$\alpha$ sequence in the simulations presented by these authors. In the Milky Way, the highest metallicity stars are found in the inner disc (R$<$6~kpc) and are the endpoint of a two-stage evolution, the first stage being the thick disc formation. This is also the case in our simulations, when such high-metallicity stars exist (model 1 and 2). The pathway to the formation of high-metallicity stars in \cite{2018MNRAS.477.5072M} is different: they are the typical end product of a low-$\alpha$ evolution that occurs in the outer regions of these simulated discs. Finally, the formation of two distinct sequences in \cite{2018MNRAS.477.5072M} seems to hinge on two circumstantial conditions: the high-$\alpha$ sequence on an early and rapid accretion of gas, which seems to be a rare event in the their simulations of MW-type galaxies, and the later accretion of gas due to the merger of a gas-rich galaxy. The consequence is that the authors conclude the formation of both sequences to be a rare event. We shall see in the next section that, on the contrary, we find that it must be relatively widespread.

\begin{figure*}
\begin{center}
\includegraphics[width=0.8\hsize]{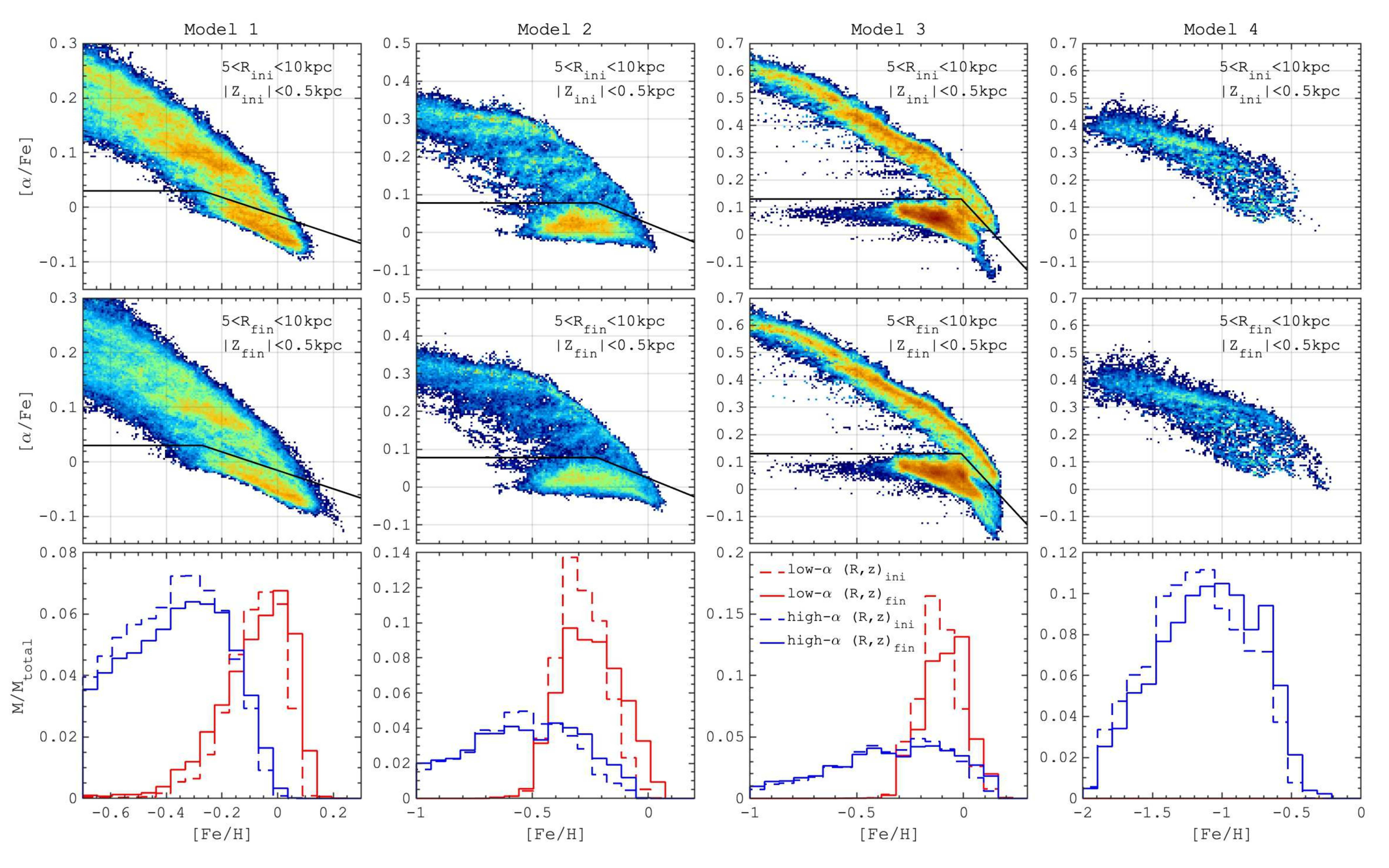}
\caption{{\it Top row}: $\aFe-\FeH$ distribution of stars {\it formed} in the middle part of the disc~($\rm 5<R_{ini}<10$~kpc)  near the galactic plane~($\rm  |Z_{ini}|<0.5$~kpc). {\it Middle row:} $\aFe-\FeH$ relation for stars {\it located at the end of simulation} in the middle part of the disc~($\rm 5<R_{fin}<10$~kpc) near the galactic plane~($\rm  |Z_{fin}|<0.5$~kpc). {\it Bottom row:} metallicity distribution for both high- and low-$\alpha$ sequences for stars formed and located at the end in the same region. Black lines in the first two rows separate high- and low-$\alpha$ sequences. Comparison of two first rows demonstrates that stellar radial migration leads to a weak blurring of the $\aFe-\FeH$ relation. However, the global structure of high- and low-$\alpha$ sequences does not depend on radial migration. Migration of stars from the innermost region leads to a slight shift of the $\FeH$ distributions towards higher values for both thick and thin discs, but the fraction of these metal-rich stars does not exceed $1-2\%$, being comparable to the estimates for the Milky Way.}\label{fig::migration3}
\end{center}
\end{figure*}

\cite{2020MNRAS.491.5435B} analysed the NIHAO-UHD cosmological zoom simulations, with results that are similar to ours. In particular, the low- and high-$\alpha$ sequences show clear bimodality, and the ridge defining the low-$\alpha$ sequence varies towards decreasing metallicity as a function of radius. \cite{2020MNRAS.491.5435B} reached the same conclusion as \cite{2018MNRAS.477.5072M} by saying that the formation of low-$\alpha$ sequence is linked to gas-rich mergers. The link is obvious in some simulated galaxies, for instance the galaxy designated {\it g7.55e11}, which shows a spread in metallicity after the merger, but much less in some others. Another example, {\it g6.96e11} apparently suffers  the greatest number of mergers, with no obvious bimodality. The most massive galaxy ({\it g2.79e12}) presents patterns very similar to those of our model 1. It is difficult to think that the two merger events linked to this galaxy (see Fig.~7 in \cite{2020MNRAS.491.5435B}), at times $\sim 3.5$ and $10$~Gyr, are directly linked to the low-alpha sequence, which stars start to form at $7$~Gyr, even if the gas provided by the mergers probably fuels the star formation activity in the disc. Ultimately, these results raise the question of the origin of the gas that settles to form the low-$\alpha$ sequence. Although the aforementioned studies emphasize the role of gas-rich mergers we suggest that a gradual (re)-accretion of gas from the halo generically give rise to a low-$\alpha$ sequence of a thin disc.

\begin{figure*}
\begin{center}
\includegraphics[width=0.8\hsize]{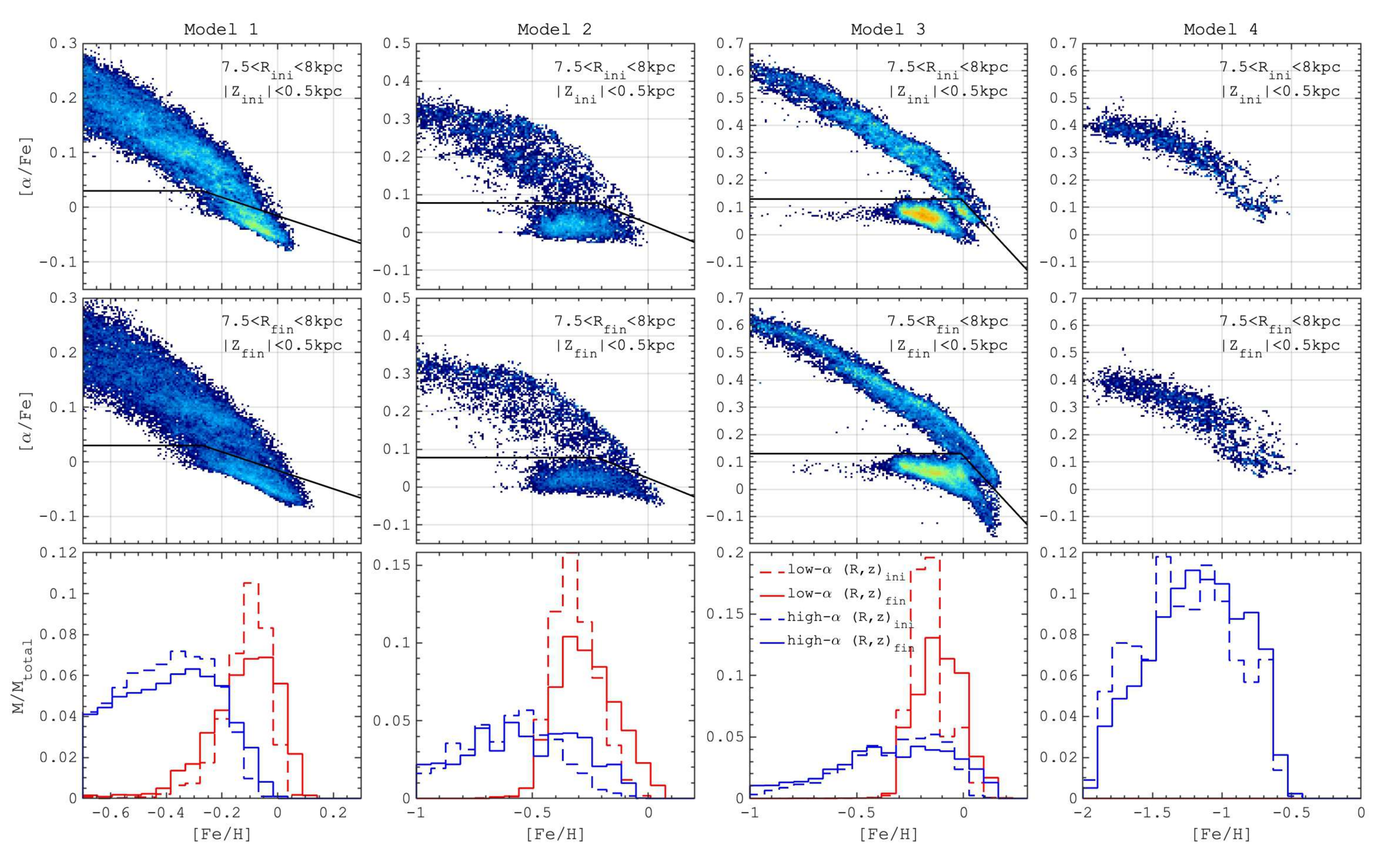}
\caption{Same as previous figure, but for the distance range 7.5 to 8~kpc.}\label{fig::migration4}
\end{center}
\end{figure*}

\subsection{How common are these patterns?}
In our simulations, a high-$\alpha$ sequence is always present at $\FeH>-1$ and is generated as a consequence of the intense star formation that occurs in the first Gyrs, reaching values of the order or higher than $\sim$10~\Msunyr for a few Gyrs. How widespread these high-$\alpha$ sequences are among galaxies will of course depend on the fraction of galaxies that start their evolution with a substantial amount of gas in their inner parts. \cite{2018MNRAS.477.5072M} advocate that this number is probably small, because galaxies with only a low-alpha sequence are much more typical in MW-like galaxies. This, however, is a potential problem, because 
MW-type galaxies are known to have a rapid mass growth, building about 50\% of their stellar mass before z$\sim1.5$~\citep[see, e.g.,][]{2013ApJ...771L..35V,2015ApJ...803...26P,2015ApJ...805...34M}, implying typical SFR of about $>$ 10 \Msunyr for a few Gyrs. 
Hence, unlike what was found by \cite{2018MNRAS.477.5072M}, we would expect high-$\alpha$ sequences to be a common feature of MW-like galaxies. This is also supported by the widespread of thick discs in the local universe \citep{2006AJ....131..226Y,2011ApJ...741...28C}. A possible solution to this apparent contradiction is the criterion adopted by \cite{2018MNRAS.477.5072M} to classify the stars in low and high-$\alpha$ stars, which is illustrated on their Fig. 5. The limit they choose to separate the two groups is relatively high in the \FeH-\aFe plane and seems to be a very conservative definition of the high-$\alpha$ stars, leading them to classify stars as "low-$\alpha$" even when they have \aFe above 0.4 or 0.5. Another equally valid definition of high-$\alpha$ stars could be that \aFe$>$0.4, which would include much more objects, and probably lead to different conclusions. Another possible explanation comes from the rather atypical SFHs of the EAGLE Milky Way-like galaxies, illustrated by the figure~5 in \cite{2018MNRAS.477.5072M}, and which barely reach intensities above 10~\Msunyr, and for a limited amount of time (much less than a Gyr in the galaxy labelled 'B'). This is not what is inferred from either observations \citep{2013ApJ...771L..35V, 2015ApJ...803...26P,2015ApJ...805...34M} or other simulations~\citep{2020MNRAS.491.5435B}. 

According to \cite{2018MNRAS.474.3629G}, \cite{2018MNRAS.477.5072M} and \cite{2020MNRAS.491.5435B}, 
the presence of bimodality is related to separate episodes of gas accretion, a significant part of the gas fueling the low-$\alpha$ sequence being related to gas-rich mergers. However, we expect that the building of discs with exponential profiles in dark matter halos will generically give rise to low-star formation efficiency, independently of any particular gas accretion history. A low-star formation efficiency is observed in outer regions of discs~\citep[see, e.g.,][]{2010AJ....140.1194B}.
These discs will give rise to the prolonged star formation history of gradually decreasing intensity outwards, driven by the decreasing gas density. Therefore, we should expect bimodality to be a ubiquitous feature in Milky Way-like galaxies. In some cases, however, the stage of prolonged star formation is never reached, most of the gas being consumed before the galaxy forms a substantial amount of stars on the low-$\alpha$ sequence. This is illustrated by the model~4, which may resemble what is expected from S0 galaxies.

\subsection{Radial migration}
Radial migration is believed to be responsible for a number of observed chemical abundance patterns in the Milky Way~\citep{2009MNRAS.399.1145S} and in the nearby galaxies~\citep[see, e.g.,][]{2016ApJ...830L..40S}. However, as we discussed in Section~\ref{sec::migration} the current structure of the Milky Way makes a spiral arms-driven migration possible only beyond the Solar radius because of its slow pattern speed and the corotation location in the outer disc. Alternatively, if we assume that spiral arms rotated much faster at the early epochs when most of the thin disc stars were not formed yet, spiral arms-induced churning could have had more impact on thick disc stars. However, once the thick disc stars formed as a dynamically hot structure, its stars were unlikely to be trapped by the spiral arm corotation due to the low fraction of stars on nearly circular orbits. Alternatively, the Milky Way spirals can be corotating structures making possible radial migration everywhere along them~\citep{2012MNRAS.426..167G, 2012MNRAS.421.1529G}. Nevertheless, this assumption is not yet supported by the maser sources kinematics, which is in favour of a rigid rotation of the spiral structure~\citep{2009ApJ...700..137R,2014ApJ...783..130R}, and a similar result was obtained using Gaia DR~2 data~\citep{2020A&A...634L...8K,2020arXiv200301132E}. Recent Gaia DR 2 analysis also suggests that the Milky Way bar can cause radial migration during its gradual slowdown~\citep{2019arXiv191204304C}, but even in this case the amount stars migrating outwards is rather modest and most of the stars~($90\%$) do not churn out more than $2-4$~kpc over the last $\approx 8$~Gyr, in agreement with some independent estimates~\citep{2018ApJ...865...96F,2018MNRAS.481.1645M}. Note, however, that both mechanisms of radial migration~(spiral arms scattering or bar-slowdown) are sufficient to explain the presence of high-metallicity stars~($\FeH>0.2$) in the Solar vicinity migrated from the innermost galaxy~\citep{2013MNRAS.436.1479K,2013A&A...558A...9M,2020A&A...638A.144K}.

\section{Summary}\label{sec::conclusions}
In this paper we present a set of self-consistent chemodynamical Milky Way-type galaxy formation simulations without mergers. 
The simulations show two-phases star formation histories that are not an input of the model but are driven by the two regimes of the gas infall that naturally arise from the pre-existing extended hot halo. The first infall is the result of a rapid collapse of the primordial gas which is halted~(in $1-3$~Gyr) by the intense stellar feedback leading to substantial removal of the gas from the galactic disc. Most of the released gas roughly conserves its angular momentum, but part of the ejected material to large radii gains significant angular momentum before re-accretion, participating to the infall of metal-richer gas on a longer time-scale, and which then contributes to the thin disc formation. Despite significant pollution of the disc-halo interface, there is a certain delay between the release of metals and their participation to the chemical evolution of the stellar disc. This leads to the existence of enriched (up to solar values) warm/hot gas which is however not involved in the star formation until it cools down via galactic fountain mechanism. We show that a proto-galaxy collapsing to a disc naturally leads to the formation of $\aFe$-bimodality. Our results and conclusions can be summarized as follows:

\begin{itemize}

\item[1.] The \aFe-\FeH bimodality is observed in galaxies with two different regimes of star formation. In our simulations, this takes place as an initial burst~($10-30$~\Msunyr) of star formation followed by a subsequent quiescent star formation phase~($3-6$~\Msunyr). The high-$\alpha$ sequence is formed early on a short time scale~($2-4$~Gyr) in a thick turbulent gaseous disc, and it is represented by a compact and kinematically hot thick stellar disc. The low-$\alpha$ sequence corresponds to the stars formed in the radially extended thin stellar disc, and is made of stars formed on a longer time-scale (last 6-7 Gyr) and which evolution at a given radius piles-up at a terminal (or equilibrium) metallicity and $\alpha$ abundance. The sequence itself arises because the equilibrium metallicity and abundance vary gradually with radius, due to the decrease of the star formation efficiency with the gas density.
Thereby, we demonstrate that different physical conditions are necessary for the formation of thick and thin galactic discs. In our simulations,  the two components are formed one after the other, implying no significant overlap of ages of thick and thin disc stars, but this needs not to be always the case and must depend on how rapidly the gas density necessary for star formation is reached in the extended disc. Stars in the thick disc are older, have enhanced $\alpha$-element abundances and lower metallicity as well as hotter kinematic features. Note that, similar to what is observed in the Milky Way, in our models with $\alpha$-bimodality, thick discs represent a substantial fraction of the total stellar mass of simulated galaxies~($\approx 40-60\%$). Contrary to the recent studies however, we do not expect that bimodality is dependent on any particular accretion history. Bimodality is related to the massive and early accretion of gas that must be prevalent in Milky Way-like galaxies to build the high-$\alpha$ sequence, and to the presence of lower star formation efficiency regions that are expected in the outer regions of all stellar discs. We therefore expect this to be a common feature in Milky Way-like galaxies.
We also find in-situ formed stars with chemical abundances typical of the low-$\alpha$ sequence and typically interpreted as made of accreted stars. Therefore, our simulations suggest that in-situ stellar populations of the Milky Way could contribute to the low $\alpha$-populations at metallicities below $\lesssim-0.5$.

\item[2.] The thick disc formation phase plays a significant role in the enrichment of the CGM surrounding Milky Way-type disc galaxies. The thin disc forms from a mixture of enriched material expelled from the thick disc and accreted gas low metallicity. When a sharp transition between thick and thin disc formation phases occurs, it is imprinted in the chemical abundances in a form of a loop~(see Fig.~\ref{fig::tracks}) in the chemical evolution tracks in the \aFe-\FeH plane. A sharper and efficient star formation quenching leads to a stronger dilution of \FeH and a more prominent loop. The effect of this non-monotonic chemical evolution can somehow mimic the impact of stellar radial migration towards in the inner parts of the disc by producing thin disc stars of lower metallicity than the youngest thick disc stars.

\item[3.] Although radial migration plays a role in mixing stellar populations, in our simulations we do not find a significant impact on the global structure of the observed chemical abundance patterns. In particular, we found very little difference in the \aFe-\FeH distribution over time in the solar-like radius~($5-10$~kpc or $7.5-8$~kpc, see Fig.~\ref{fig::migration3} and Fig.~\ref{fig::migration4}) confirming that $\aFe-\FeH$ bimodality in the Milky Way is the outcome of the thin and thin disc formation process and is not produced by secular evolution. Nevertheless, the high-\FeH tail of the metallicity distribution, representing a few percent of the stars in the simulations, is explained by radial migration, has it is for stars in the solar vicinity.
Outward churning dominates in all our simulations due to the negative density gradient of the discs. Models with a bar show a more substantial redistribution of stellar orbits in the inner disc~(see models 1, 2 in Fig.~\ref{fig::migration1}) while spiral arms induced more prominent migration in the outer disc where the corotation radius is located~(see Model 4 in Fig.~\ref{fig::migration1}). 

\end{itemize}

\section*{Acknowledgements}
We thank Chris Brook for providing a constructive referee report that helped to improve the paper. PDM and MH thank the ANR (Agence Nationale de la Recherche) for its financial support through the MOD4Gaia project (ANR-15- CE31-0007, P.I.: P. Di Matteo). This work was granted access to the HPC resources of CINES under the allocation 2017-040507 (PI : P. Di Matteo) made by GENCI. Numerical simulations were partially carried by using the equipment of the shared research facilities of HPC computing resources at Lomonosov Moscow State University supported by the project RFMEFI62117X0011. SN and EV acknowledge support from Russian Science Foundation~(project no. 19-72-20089). ONS acknowledges DIM ACAV+ funding. The reported study was partially funded by RFBR and DFG according to the research project 20-52-12009. PB acknowledges support by the Chinese Academy of Sciences through the Silk Road Project at NAOC, the President's International Fellowship (PIFI) for Visiting Scientists program of CAS, the National Science Foundation of China under grant No. 11673032. This work was supported by the Deutsche Forschungsgemeinschaft
(DFG, German Research Foundation) - Project-ID 138713538 - SFB 881 ('The Milky Way System'), by the Volkswagen Foundation under the Trilateral Partnerships grants No. 90411 and 97778. The work of PB was supported under the special program of the NRF of Ukraine 'Leading and Young Scientists Research Support' - ``Astrophysical Relativistic Galactic Objects (ARGO): life cycle of active nucleus'',  No. 2020.02/0346.

\bibliographystyle{mnras}
\bibliography{cems_paper.bib}

\appendix

\section{Extra plots}\label{app}
In this section we provide the chemical abundance patterns~(\aFe-\FeH) at the end of simulation in models 2, 3 and 4, similar to the Fig.~\ref{fig::hayden1} for model 1. In Fig.~\ref{fig::cem234} we show the evolution of \FeH, \aH, \aFe and the local star formation rate as a function of time in models 2, 3, 4, similar to Fig.~\ref{fig::cem} for model~1.

\begin{figure}
\begin{center}
\includegraphics[width=0.8\hsize]{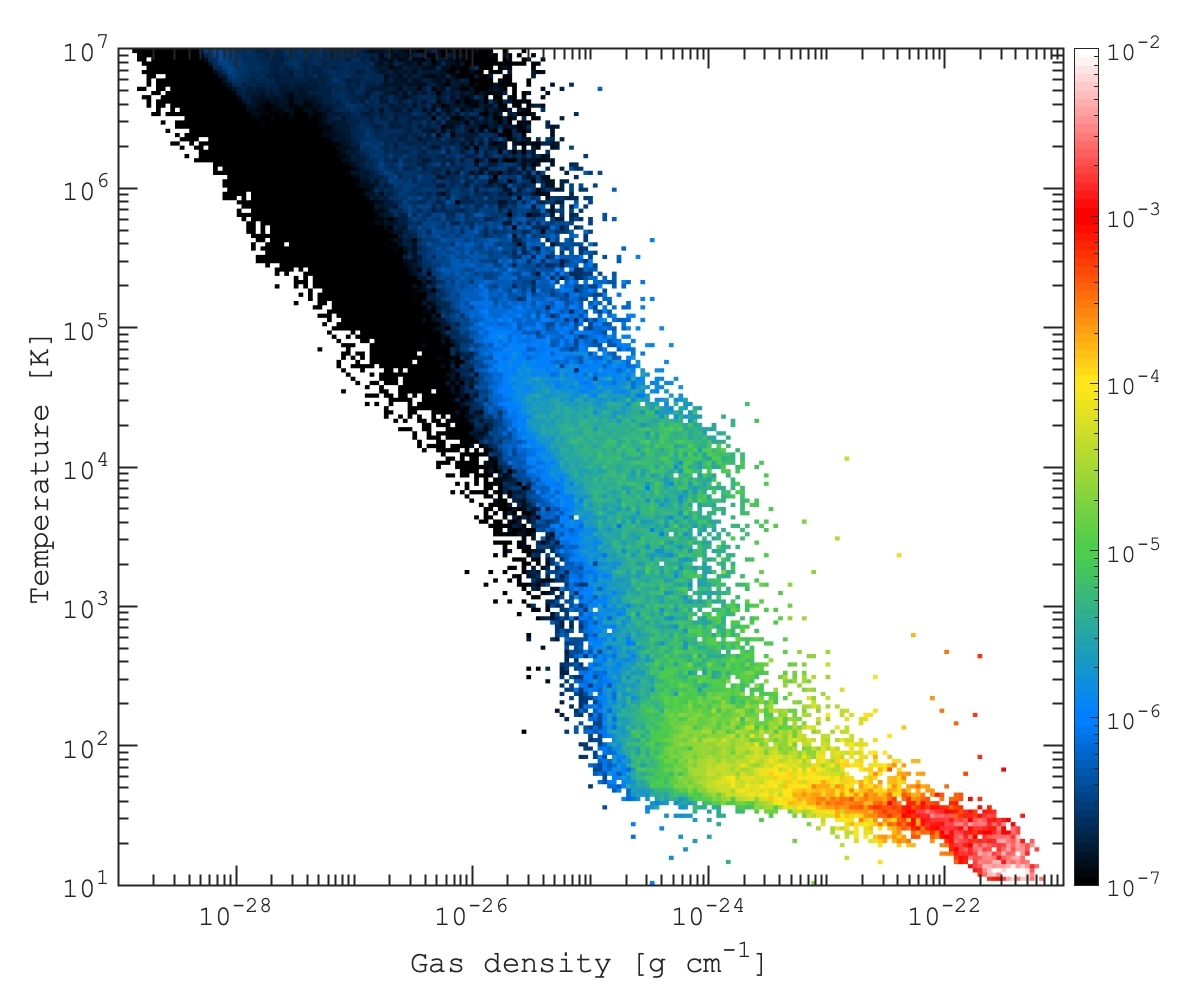}
\caption{ Example of two-dimensional probability distribution function of density and temperature for the gas in Model 1. Colors represent the total gas mass in each two-dimensional bin.}\label{fig::ISM}
\end{center}
\end{figure}

\begin{figure}
\begin{center}
\includegraphics[width=1\hsize]{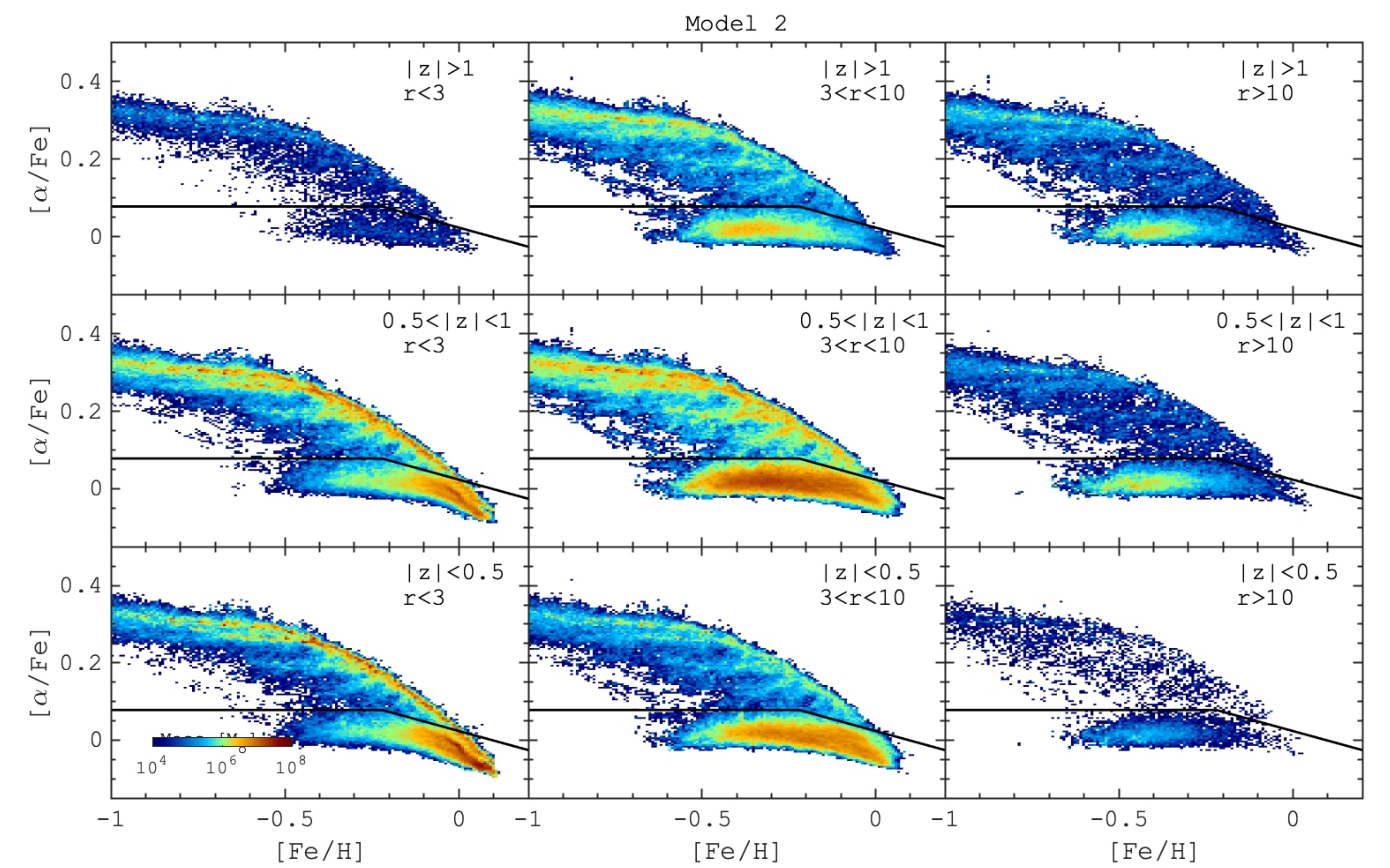}
\includegraphics[width=1\hsize]{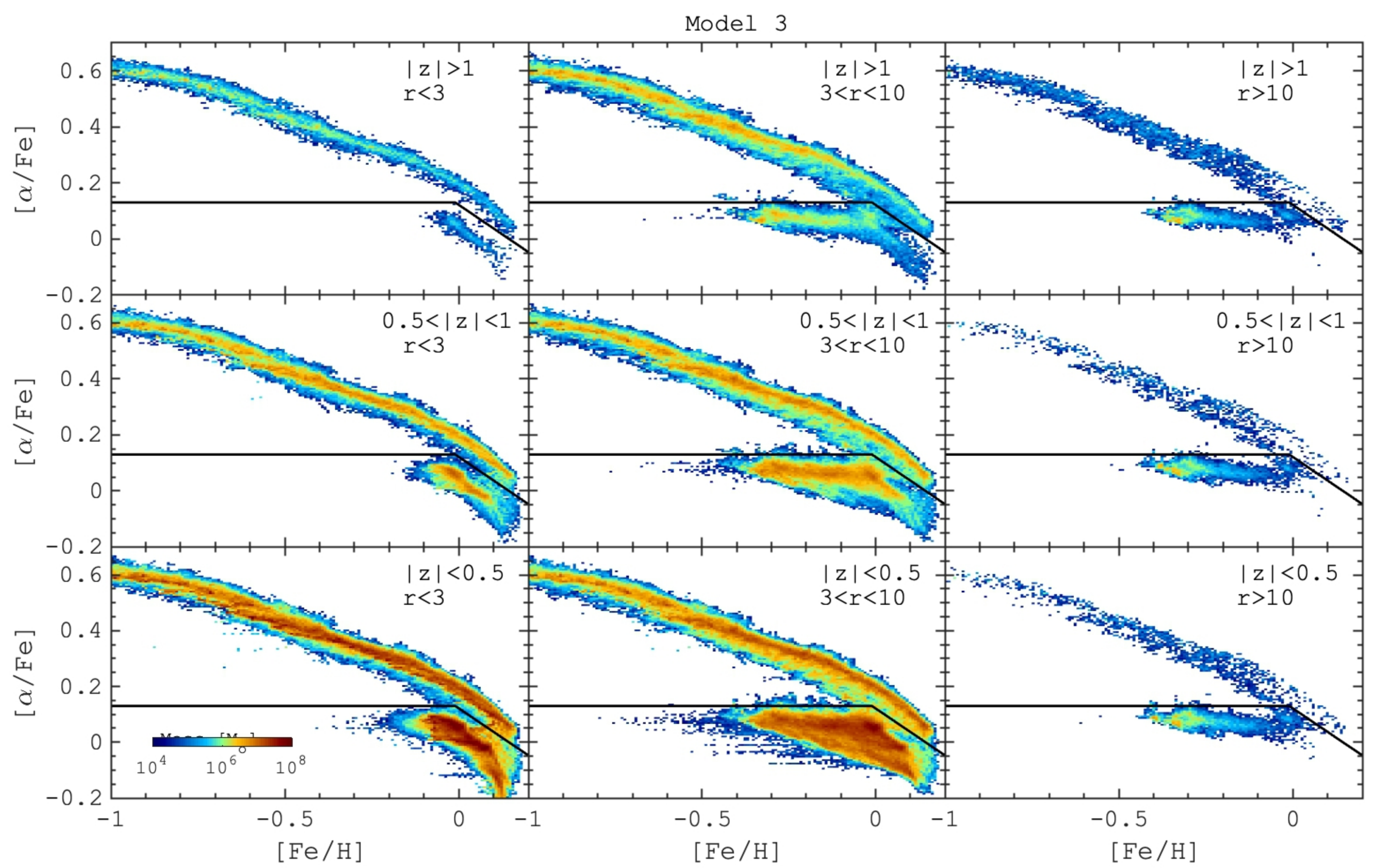}
\includegraphics[width=1\hsize]{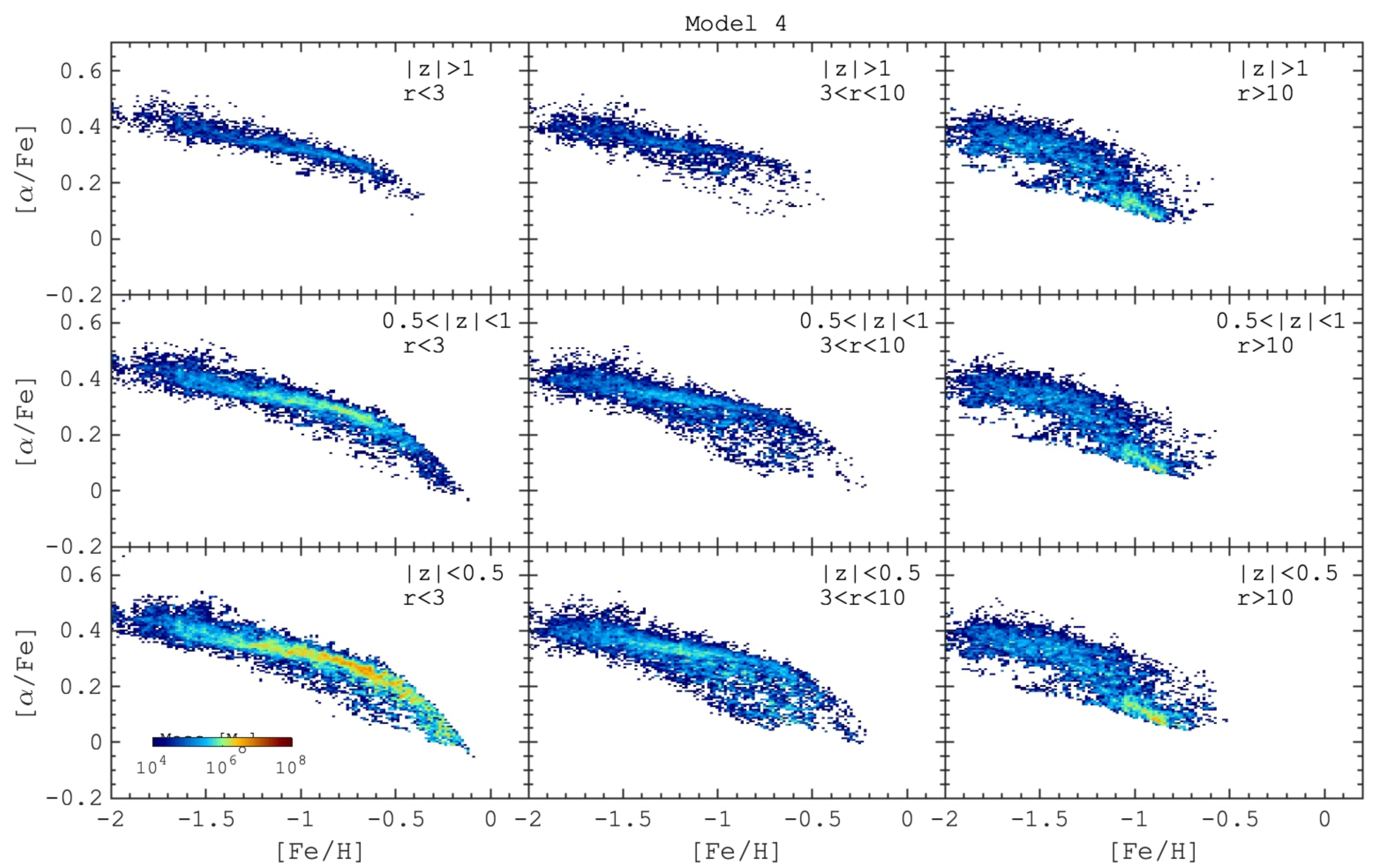}
\caption{Same as in Fig.~\ref{fig::hayden1} but for Models 2, 3 and 4.}\label{fig::hayden234}
\end{center}
\end{figure}

\begin{figure}
\begin{center}
\includegraphics[width=1\hsize]{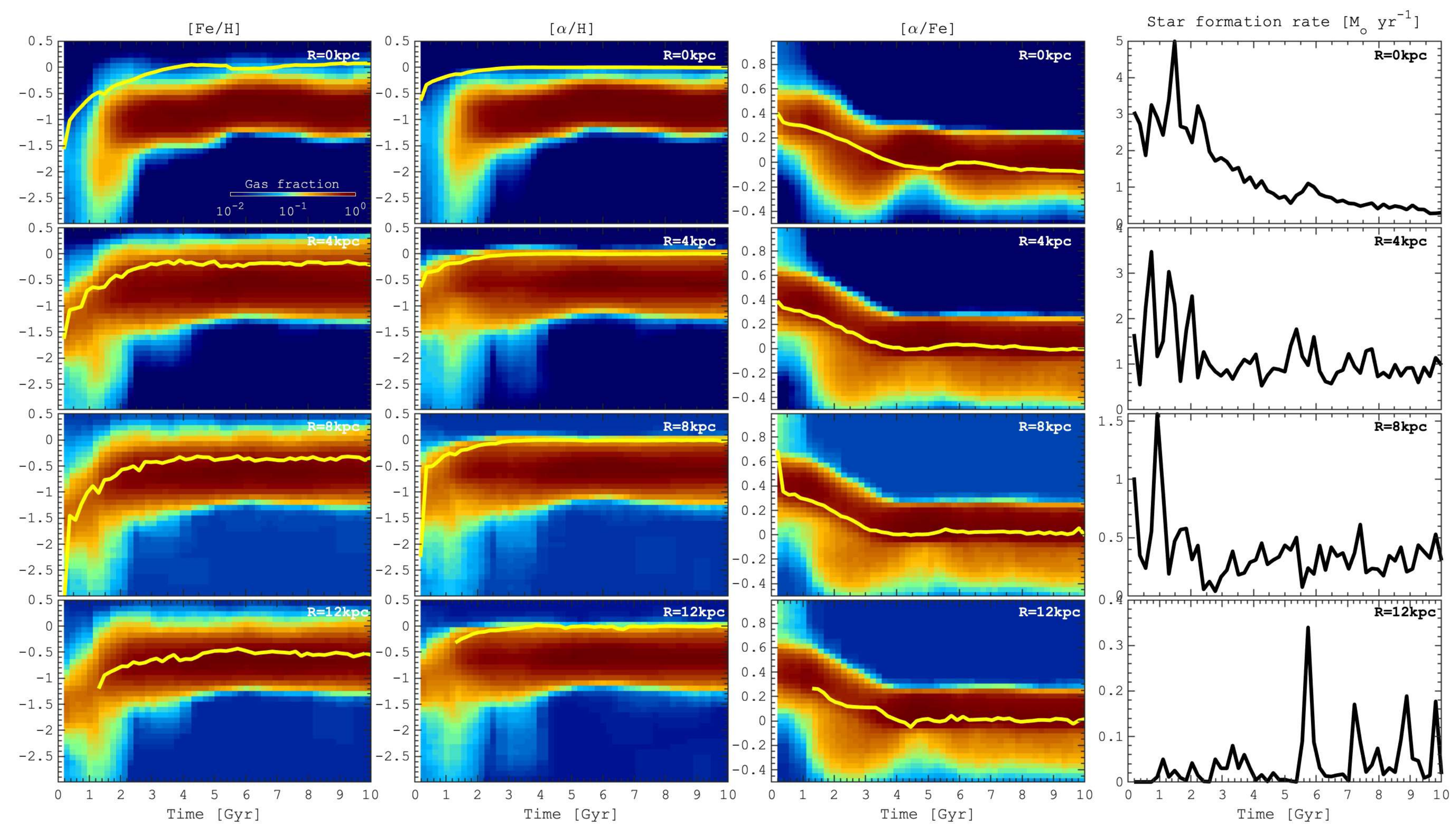}
\includegraphics[width=1\hsize]{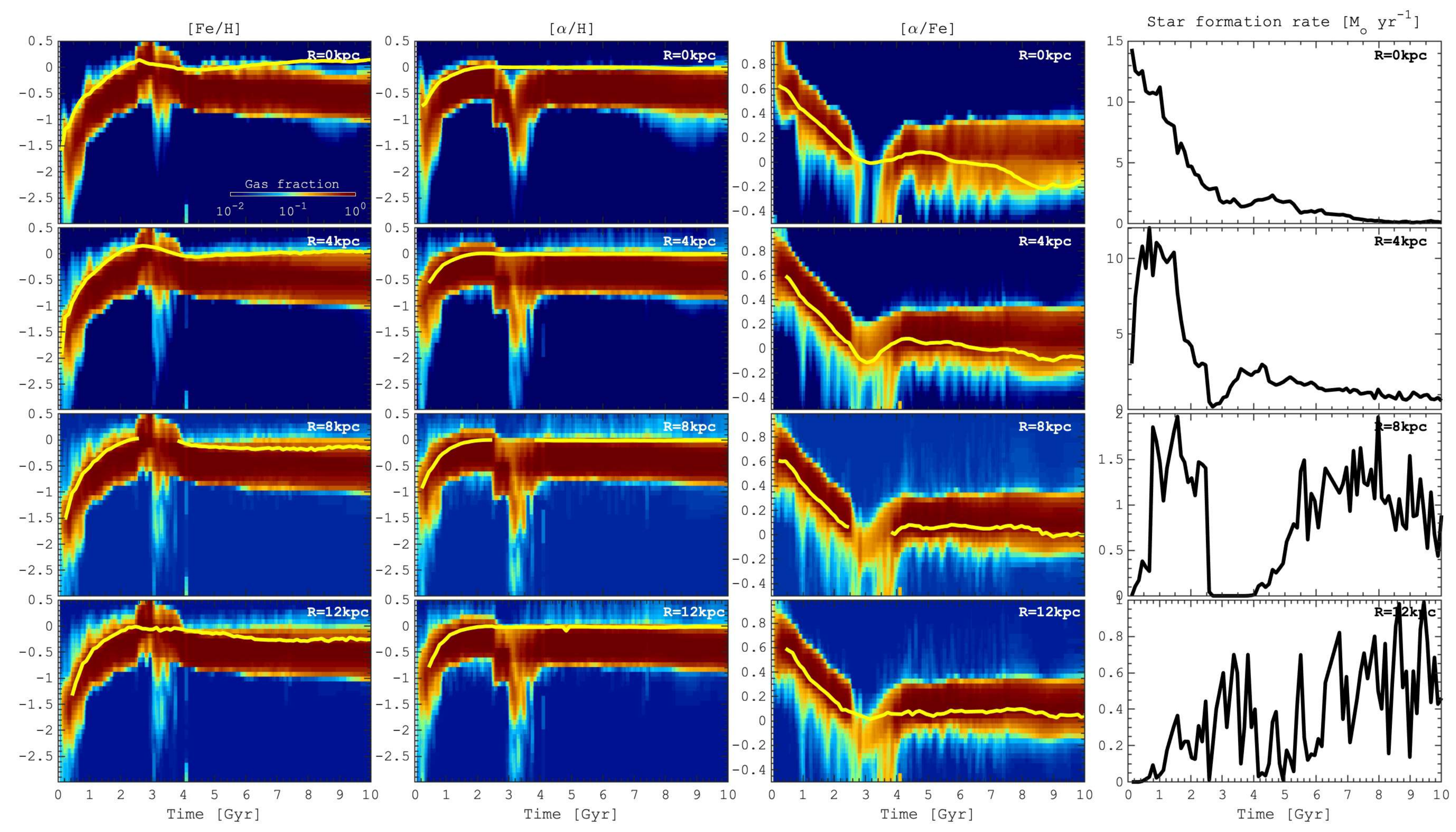}
\includegraphics[width=1\hsize]{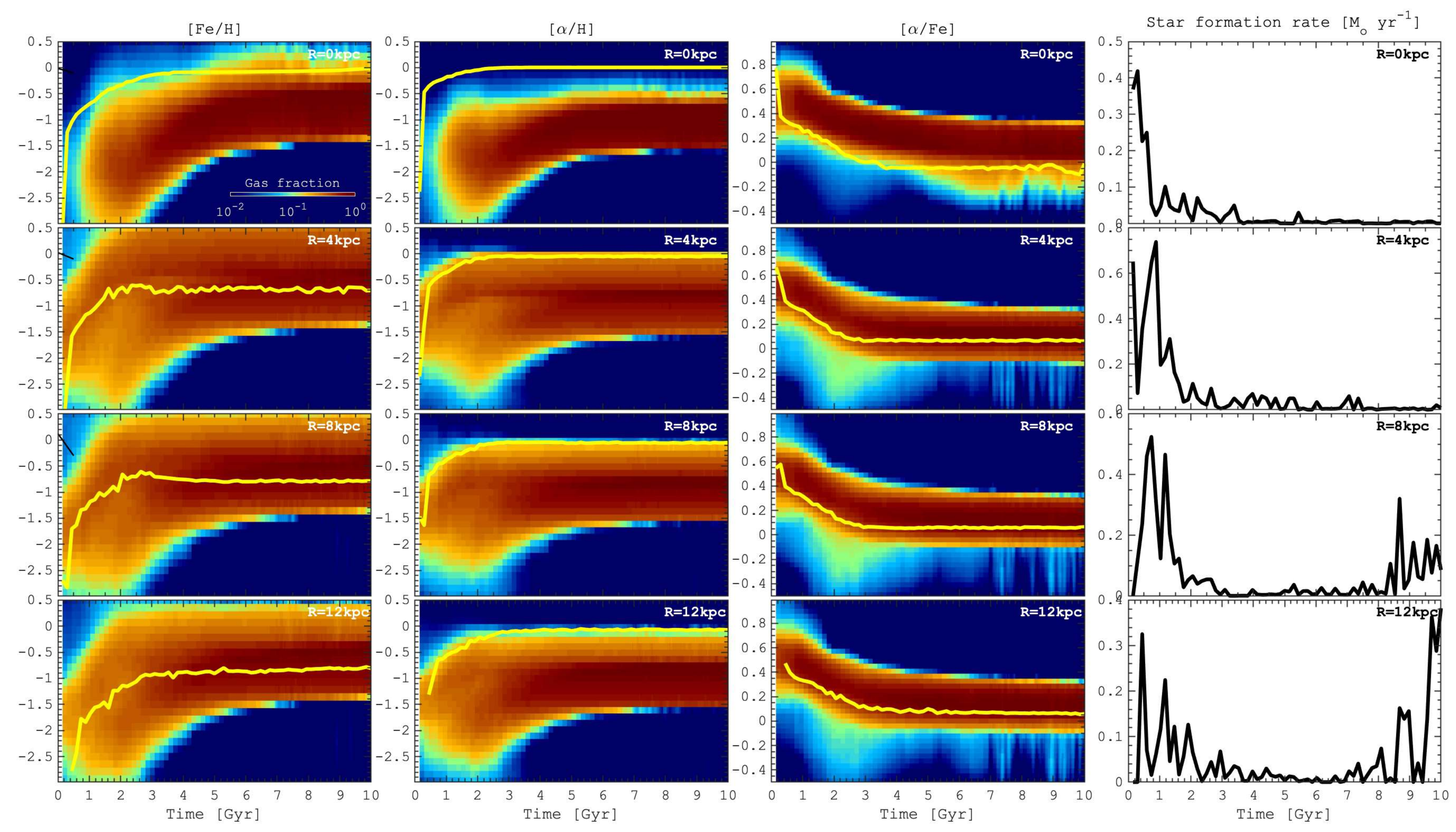}
\caption{Same as in Fig.~\ref{fig::cem} but for Models 2, 3 and 4.}\label{fig::cem234}
\end{center}
\end{figure}

\bsp	
\label{lastpage}
\end{document}